\documentclass[aps,prd,twocolumn,showpacs,superscriptaddress,groupedaddress]{revtex4}

\usepackage{dcolumn}   
\usepackage{bm}        

\usepackage{amsmath}
\usepackage{amsfonts}
\usepackage{amssymb}
\usepackage{makeidx}
\usepackage{graphicx}
\usepackage{anysize}

\newcommand{\ba}{\begin{eqnarray}}
\newcommand{\ea}{\end{eqnarray}}
\newcommand{\be}{\begin{equation}}
\newcommand{\ee}{\end{equation}}

\usepackage[dvips]{color}

\begin{document}
\title{Extensions of Superscaling from Relativistic \\ Mean Field Theory:
the SuSAv2 Model}

\author{R.~Gonz\'alez-Jim\'enez}
\affiliation{Departamento de F\'isica At\'omica, Molecular y Nuclear, Universidad de Sevilla, 41080 Sevilla, Spain}

\author{G.D.~Megias}
\affiliation{Departamento de F\'isica At\'omica, Molecular y Nuclear, Universidad de Sevilla, 41080 Sevilla, Spain}

\author{M.B.~Barbaro}
\affiliation{Dipartimento di Fisica, Universit\`a di Torino and INFN, Sezione di Torino, 10125 Torino, Italy}

\author{J.A.~Caballero}
\affiliation{Departamento de F\'isica At\'omica, Molecular y Nuclear, Universidad de Sevilla, 41080 Sevilla, Spain}

\author{T.W.~Donnelly}
\affiliation{Center for Theoretical Physics, Laboratory for Nuclear Science and Department of Physics,
Massachusetts Institute of Technology, Cambridge, Massachusetts 02139, USA}

\date{\today}

\begin{abstract}
We present a systematic analysis of the quasielastic scaling functions
computed within the Relativistic Mean Field (RMF) Theory and we propose
an extension of the SuperScaling Approach (SuSA) model based on these
results.
The main aim of this work is to develop a realistic and accurate
phenomenological model (SuSAv2), which incorporates the
different RMF effects in the longitudinal and transverse nuclear responses,
as well as in the isovector and isoscalar channels. This provides a complete
set of reference scaling functions to describe in a consistent way
both $(e,e')$ processes and the neutrino/antineutrino-nucleus reactions in
the quasielastic region.
A comparison of the model predictions with electron and neutrino
scattering data is presented.
\end{abstract}

\maketitle

\tableofcontents

\section{Introduction}

Scaling is a phenomenon observed in several areas of
Physics~\cite{West75}. It occurs when a particle interacts with a
Many-body system in such a way that energy $\omega$ and momentum $q$
are transferred only to individual constituents of the complex
system. In the particular case of quasielastic (QE) scattering of
electrons from nuclei, in most of the models based on the Impulse
Approximation (IA), the inclusive $(e,e')$ cross section can be
written approximately as a single-nucleon cross section times a
specific function of $(q,\omega)$. Scaling occurs when, in the limit
of high momentum transfers, that specific function scales, becoming
dependent on only a single quantity, namely, the {\it scaling
variable} $\psi$. This quantity, whose definition is discussed
later, is in turn a function of $q$ and $\omega$:
$\psi=\psi(q,\omega)$. The function that results once the
single-nucleon cross section has been divided out is called the {\it
scaling function} $f=f(q,\psi)$. In other words, to the extent that
at high $q$ this function depends on $\psi$, but not on $q$, one
says that $\psi$-scaling occurs.

The study of the scaling function can shed light on the dynamics of
the nuclear system. Indeed, within some specific approaches, the
scaling function is related to the momentum distribution of the
nucleons in the nucleus (or, more generally, with the spectral
function)~\cite{Caballero10b,Antonov11}.

When studying $(e,e')$ processes it is useful to introduce the following concepts:
\begin{itemize}
 \item \textbf{\textit{Scaling} of first kind}:
 This is what is discussed above: it is satisfied when the scaling function 
 does not explicitly depend on the transferred momentum, but only on $\psi$ 
 including its implicit dependence on $q$ and $\omega$.
 \item \textbf{\textit{Scaling} of second kind}:
 It is observed when the scaling function is independent of the nuclear species.
 \item \textbf{\textit{Scaling} of zeroth kind}:
 It occurs when the scaling functions linked to the different channels that 
 make up the cross section, longitudinal (L) and transverse (T), are equal. 
 For example, when considering inclusive electron scattering,
 zeroth-kind scaling means that the electromagnetic (EM) scaling functions 
 satisfy $f=f_L=f_T$, where $f$ represents the total EM scaling function and 
 $f_{L,T}$ are the EM longitudinal and transverse ones.
\item \textbf{\textit{Superscaling}}:
 Finally, when scaling of both the first and second kinds occurs simultaneously
 one has superscaling~\cite{Donnelly99a,Donnelly99b}.

\end{itemize}

The Relativistic Fermi gas (RFG) model, in spite of its simplicity,
provides a completely relativistic description of the QE process and
allows for fully analytical
expressions~\cite{Alberico88,Donnelly99b}.
Additionally, the RFG model satisfies exactly all of the kinds of scaling introduced above.
Following the formalism of the
\cite{Donnelly99a,Donnelly99b,Maieron02}, in this work we use
the RFG cross sections to build the EM scaling functions
($f_{L,T}$). The general procedure used to define scaling functions
consists in constructing the inclusive cross section, or response
functions, within a particular model (or data) and then dividing
them by the corresponding single-nucleon quantity computed
within the RFG model. The explicit expressions for the RFG single-nucleon cross
section and response functions are given in
Appendix~\ref{apendiceScaling}.


In previous
work~\cite{Day90,Jourdan96,Donnelly99a,Donnelly99b,Maieron02} a
large body of $(e,e')$ cross section data were analyzed within
this scaling formalism. The results show that first-kind
scaling works reasonably well in the region $\omega<\omega_{QEP}$
($\omega_{QEP}$ being the transferred energy corresponding to the
quasielastic peak), while second-kind scaling
is excellent in the same region of $\omega$. In contrast, when
$\omega>\omega_{QEP}$ both first- and second-kind scaling are
seen to be violated.

In \cite{Donnelly99b,Maieron02} scaling was studied by analyzing
experimental data for the individual EM longitudinal ($R_L$) and
transverse ($R_T$) responses. Those studies concluded that $f_L$
superscales approximately throughout the region of the quasielastic
peak, while $f_T$ only superscales in the region
$\omega<\omega_{QEP}$, and clearly does not for
$\omega>\omega_{QEP}$.
The scaling violation in the transverse response at high $\omega$ occurs because in
that range of the spectrum other
non-QE processes such as meson production and resonance excitation, at high
excitation energies going over into deep inelastic scattering, and excitation of
np-nh states induced by meson-exchange currents
are known to be of importance for a correct
interpretation of the scattering process.

Exploiting the superscaling property exhibited by the longitudinal
data, in \cite{Maieron02} the ``{\it experimental longitudinal
scaling function}'', namely, $f_{L,\text{exp}}^{ee'}$, extracted
from the analysis of the longitudinal
response for several nuclear species and kinematical situations, was
presented.
However, due to the non-QE contributions discussed above, the extraction of an experimental transverse
scaling function, $f^{ee'}_{T,\text{exp}}$, has not been systematically performed
to date.
%
%
Nevertheless, in spite of the difficulty of analyzing the transverse scaling
function, preliminary studies~\cite{Maieron09}, based on the modeling
of the QE longitudinal response and contributions from non-QE
channels, have provided some evidence that the scaling of zeroth kind is
not fully satisfied by data. In particular, these studies find
$f^{ee'}_{T,\text{exp}}>f^{ee'}_{L,\text{exp}}$, a point that will be discussed in more detail later.


The SuperScaling Approach (SuSA) is based on the scaling
properties of the longitudinal response extracted from $(e,e')$ data to predict
Charge Changing (CC) QE neutrino- and antineutrino-nucleus cross
sections~\cite{Amaro05a}, namely $(\nu_l,l^-)$ and $({\bar\nu}_l,l^+)$. Thus, SuSA is based on
the hypothesis that the neutrino cross section scales as does the
electron scattering cross section. This feature is observed in most
of the models based on IA (see, for instance,
\cite{Amaro05b,Caballero06,Caballero05}).
SuSA uses the experimental scaling function $f_{L,\text{exp}}^{ee'}$
as a universal scaling function and then builds the different
nuclear responses by multiplying it by the corresponding
single-nucleon responses. However, notice that the extraction of
$f_{L,\text{exp}}^{ee'}$ entails the analysis of the
longitudinal $(e,e')$ (isoscalar + isovector) nuclear response. In
contrast, CC neutrino-nucleus reactions involve only isovector
couplings and are mainly dominated by purely transverse responses
($T_{VV}+T_{AA}$ and $T'_{VA}$, the indices $V$ and $A$ referring to
the vector and axial components of the weak hadronic current). Thus,
one could question the validity of the SuperScaling Approach.
This issue was studied in \cite{Caballero07} by analyzing the
scaling functions of the Relativistic Mean Field (RMF) model (see
below). There, it was found that, contrary to what one might expect,
the $(e,e')$ longitudinal scaling function agrees with the total
$(\nu_l,l^-)$ one (which is mainly transverse) much better than does
the transverse scaling function from $(e,e')$. This result is
explained by the different roles played by the isovector and
isoscalar nucleon form factors in each process (see
\cite{Caballero07} for details).


Within the RMF model the bound and scattered nucleon wave functions
are solutions of the Dirac-Hartree equation in the presence of
energy-independent real scalar (attractive) and vector (repulsive)
potentials. Since the same relativistic potential is used to
describe the initial and final nucleon states, the model is shown to
preserve the continuity equation (this is strictly true for the CC2
current operator); hence the results are almost independent of the
particular gauge selected~\cite{Caballero05,Caballero06}.
The RMF approach has achieved significant success in describing QE electron scattering data.
On the one hand, its validity has been widely proven through
comparisons with QE $(e,e')$ data (see \cite{Caballero06,Meucci09}
and Sect.~\ref{ee-results}). In this connection, an important result
is that the model reproduces surprisingly well the magnitude and
shape of $f_{L,\text{exp}}^{ee'}$, {\it i.e.,} it yields an
asymmetric longitudinal scaling function, with  more strength in the
high-$\omega$ tail, and with a maximum value ($\sim$0.6) very close
to the experimental one.
On the other hand, the model predicts $f_T^{ee'}>f_L^{ee'}$. This
violation of zeroth-kind scaling was analyzed in
~\cite{Caballero07}, where it was shown that the origin of such an
effect lies in the distortion of the lower components of the
outgoing nucleon Dirac wave function by the final-state interactions
(FSI).

However, the RMF model also presents some drawbacks. First, it
predicts a strong dependence of the scaling function on the
transferred momentum $q$, an occurrence that is hardly acceptable given the above phenomenological discussion. For increasing values of $q$
the RMF model presents: i) a strong shift of the scaling functions to
higher $\omega$ values, ii) too much enhancement of the area under the
tail of the functions, and iii) correspondingly too severe a decrease
in the maximum of the scaling functions.
All of these features will be studied in detail in Sect.~\ref{scalingRMF}.
Second, getting results with the RMF model is computationally very
expensive, especially when the model is employed to predict neutrino cross
sections where one has to fold in the flux distribution of the incident
neutrino or one has to compute totally integrated cross sections.
Hence in what follows, after correcting for the too strong
$q$-dependence of the RMF model, we shall implement the main
features of the model in a new version of the SuSA approach, called
``SuSA version 2'', or ``SuSAv2'', that makes it possible to obtain
numerical predictions to compare with data using fast codes, yet
retaining some of the basic physics of the RMF.

In summary, the main goal of this work is to extend the SuSA model,
incorporating in its formalism information from the RMF model. So we
build the new model in such a way that it reproduces the
experimental longitudinal scaling function, produces
$f_T^{ee'}>f_L^{ee'}$, takes into account the differences in the
isoscalar/isovector scaling functions and avoids the problems of the
RMF model in the region of medium and high momentum transfer.

The structure of this work is as follows: In
Sect.~\ref{scalingRMF} we present and discuss the features of the
various scaling functions in the RMF model. In
Sect.~\ref{SuSAv2definition} we define the SuSAv2 model. In
Sects.~\ref{ee-results} and \ref{nu-results} we present the SuSAv2
results for QE electron and neutrino scattering reactions,
respectively, and compared them with selected experimental
data. In Sect.~\ref{conclusions} we draw our main conclusions.
Some details on the definitions of scaling functions and on the
implementation of Pauli blocking in the SuSAv2 approach are
presented in the Appendices.

\section{RMF scaling behavior}\label{scalingRMF}

In this section we present a systematic analysis of the scaling
functions computed with the Relativistic Mean Field (RMF) and the
Relativistic Plane Wave Impulse Approximation (RPWIA). Both models
are based on the relativistic impulse approximation (RIA) and
provide a completely relativistic description of the scattering
process. The bound state Dirac-spinors are the same in both models
and correspond to the solutions of the Dirac equation with scalar and vector potentials. The two models differ in the
treatment of the final state: the RPWIA describes the outgoing
nucleon as a relativistic plane wave while the RMF model accounts
for the FSI between the outgoing nucleon and the residual nucleus
using the same mean field as used for the bound nucleon.

In this work we analyze the scaling functions involved in the
$(e,e')$, $(\nu,\mu^-)$ and $(\bar{\nu},\mu^+)$ reactions as
functions of $q$. Because there exists a great number of $(e,e')$
and $(\nu_l,l)$ experimental data for $^{12}$C, in this work we have
chosen it as reference target nucleus.

We first split all different response functions by isolating the
isoscalar ($T=0$) and isovector ($T=1$) contributions in electron
scattering, and the Vector and Axial contributions for neutrino and
antineutrino induced reactions: VV (vector-vector), AA
(axial-axial), VA (vector-axial). This strategy will allow us to
extract clear information on how the FSI affect the different
sectors of the nuclear current. Furthermore, it will make it easier to
explore the relationships between the different responses linked to
$(e,e')$, $(\nu,\mu^-)$ and $(\bar{\nu},\mu^+)$ reactions.

The $(e,e')$ inclusive cross section, double differential with respect to the
electron scattering angle $\Omega_e$ and the transferred energy
$\omega$, is defined in terms of two response functions corresponding
to the longitudinal, $R_L$, and transverse, $R_T$, channels ($L$ and
$T$ refer to the direction of the transferred momentum, ${\bf q}$).
It reads \ba \frac{d^2\sigma}{d\Omega_e d\omega} = \sigma_{Mott}
\left(v_LR_L + v_T R_T\right)\,, \ea where $\sigma_{Mott}$ is the
Mott cross section and the $v$'s are kinematical factors that
involve leptonic variables (see~\cite{Day90} for explicit
expressions). Assuming charge symmetry, these two channels can be
decomposed as a sum of the isoscalar ($T=0$) and isovector ($T=1$)
contributions. In terms of the scaling functions (see
~\cite{Donnelly99b}) the nuclear responses are: \ba
 R_{L,T}^{ee'}(q,\omega) &=& \frac{1}{k_F}
    \left[f_{L,T}^{T=1,ee'}(\psi') G_{L,T}^{T=1}(q,\omega)\right.\nonumber\\
           &+& \left. f_{L,T}^{T=0,ee'}(\psi')G_{L,T}^{T=0}(q,\omega)\right]\,.
           \label{RLTee}
\ea

Similarly, the charge-changing muon-neutrino (antineutrino) cross section is~\cite{Amaro05a}:
\ba
\frac{d^2\sigma}{d\Omega_\mu d\varepsilon_\mu} &=& \sigma_0 \left(
\hat{V}_LR^{VV}_L + \hat{V}_{CC}R^{AA}_{CC} + 2\hat{V}_{CL}R^{AA}_{CL}\right.\nonumber\\
&+& \left.\hat{V}_{LL}R^{AA}_{LL} + \hat{V}_T R_T +
\chi\hat{V}_{T'}R_{T'}\right)\,, \ea where $\Omega_\mu$ and
$\epsilon_\mu$ are the scattering angle and energy of the outgoing
muon, $\chi=+$ for neutrino-induced reactions and $\chi=-$ for
antineutrino ones, $\sigma_0$ is the equivalent to the Mott cross
section in CC neutrino reactions and the $\hat{V}$'s are leptonic
kinematical factors (see~\cite{Amaro05a,Amaro05b} for explicit
expressions). In this case, the responses are: \ba
 R_{L}^{VV,\nu(\bar{\nu})}(q,\omega) &=& \frac{1}{k_F}
      f_{L}^{VV,\nu(\bar{\nu})}(\psi')G_{L}^{VV}(q,\omega) \label{RLVV}\\
 R_{CC}^{AA,\nu(\bar{\nu})}(q,\omega) &=& \frac{1}{k_F}
      f_{CC}^{AA,\nu(\bar{\nu})}(\psi')G_{CC}^{AA}(q,\omega) \\
 R_{CL}^{AA,\nu(\bar{\nu})}(q,\omega) &=& \frac{1}{k_F}
      f_{CL}^{AA,\nu(\bar{\nu})}(\psi')G_{CL}^{AA}(q,\omega) \\
 R_{LL}^{AA,\nu(\bar{\nu})}(q,\omega) &=& \frac{1}{k_F}
      f_{LL}^{AA,\nu(\bar{\nu})}(\psi')G_{LL}^{AA}(q,\omega) \\
 R_{T}^{\nu(\bar{\nu})}(q,\omega) &=& \frac{1}{k_F}
       \left[f_{T}^{VV,\nu(\bar{\nu})}(\psi')
        G_{T}^{VV}(q,\omega)\right.\nonumber\\
 &+& \left.f_{T}^{AA,\nu(\bar{\nu})}(\psi')
    G_{T}^{AA}(q,\omega)\right]
    \label{RTnunub}\\
 R_{T'}^{\nu(\bar{\nu})}(q,\omega) &=& \frac{1}{k_F}
   f_{T'}^{VA,\nu(\bar{\nu})}(\psi') G_{T'}^{VA}(q,\omega).
   \label{RTp}
\end{eqnarray}

The $G$s in Eq.~(\ref{RLTee}) and Eqs.~(\ref{RLVV}--\ref{RTp})
are the single-nucleon responses from RFG that are defined in
Appendix~\ref{apendiceScaling}. The $f$'s are the scaling functions
which --- if scaling is fulfilled --- only depend on the scaling
variable $\psi'$, also defined in Appendix~\ref{apendiceScaling}.
The scaling variable $\psi'$ depends on $q$, $\omega$ and on the
energy shift, $E_{shift}$, which is introduced to reproduce the
position of the experimental QE peak (see
Appendix~\ref{apendiceScaling}).

In the following we examine three basic features of the scaling
functions in the RPWIA and RMF models: shape, position and height of
the peak, and the integrals of the scaling functions over
$\psi'$~\cite{Caballero10a}.

\subsection{Shape of the scaling functions}\label{shape}

\begin{figure}[htbp]
     \centering
         \includegraphics[width=0.34\textwidth,angle=270]{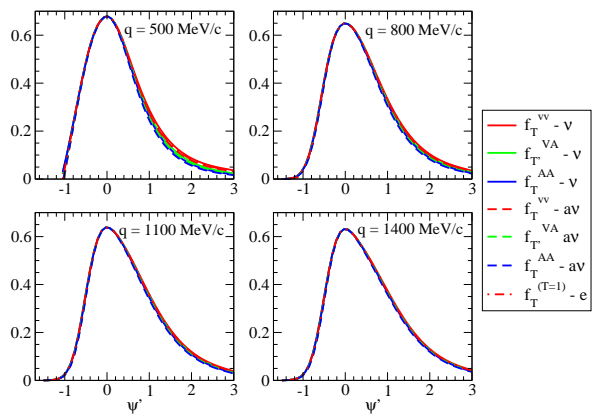}
     \caption{Transverse RMF scaling functions normalized to the maximum value corresponding to an arbitrary reference function and relocated at $\psi'=0$
     (see text for details).
     The convention used to label the different curves is as follows:
     ``$e$'' for electron-induced reactions and ``$\nu$'' (``a$\nu$'') for neutrino- (antineutrino-) induced reactions.}
     \label{fig:normalized_T}
\end{figure}
\begin{figure}[htbp]
     \centering
         \includegraphics[width=0.34\textwidth,angle=270]{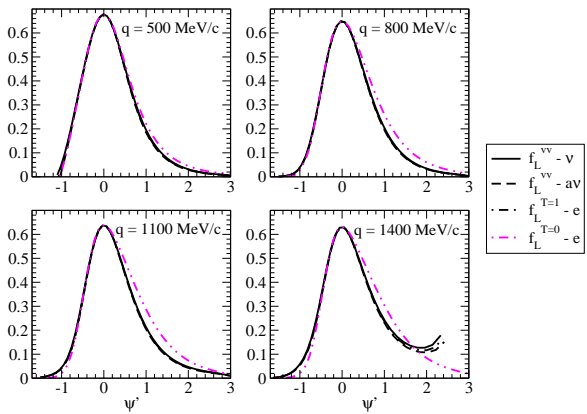}
     \caption{As in Fig.~\ref{fig:normalized_T}, but now for the longitudinal RMF scaling functions.}
     \label{fig:normalized_L}
 \end{figure}
\begin{figure}[htbp]
     \centering
         \includegraphics[width=0.34\textwidth,angle=270]{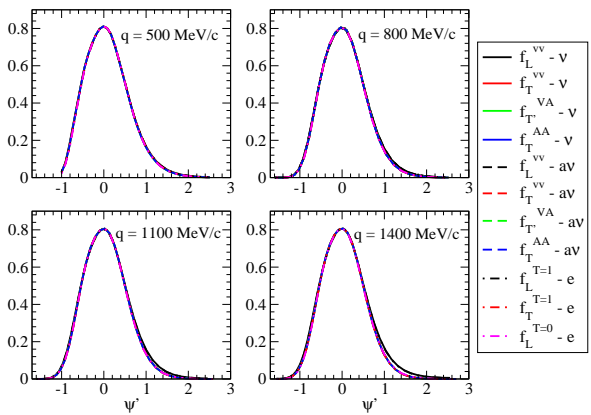}
     \caption{As in Fig.~\ref{fig:normalized_T}, but in this case the
     results correspond to RPWIA.
     Transverse and longitudinal sets are presented together.}
     \label{fig:normalized_rpwia}
 \end{figure}

The goal here is to study the shape of all scaling functions. In
Fig.~\ref{fig:normalized_T} (Fig.~\ref{fig:normalized_L}),
for different values of $q$, we
present the transverse (longitudinal)
RMF scaling functions normalized to the maximum value
corresponding to a reference function, in this case $f_T^{VV,\nu}$,
and relocated so that the maximum is at $\psi'=0$. As already
mentioned, the scaling variable $\psi'$ depends on $q$, $\omega$ and
$E_{shift}$. Thus, for each scaling function, $E_{shift}$ is taken
so that the maximum is located at $\psi'=0$. The results within the
RPWIA model are presented in Fig.~\ref{fig:normalized_rpwia}.


We do not present results of $f_{CC}^{AA}$, $f_{CL}^{AA}$,
$f_{LL}^{AA}$ for neutrino and antineutrino scattering, and
$f_T^{T=0}$ for electron scattering because they are very sensitive to small effects
due to cancellations and/or to the smallness of the
denominator ($G$ function) which appears in the definition of the
scaling function (see Appendix~\ref{apendiceScaling}).
The first three are seen to be insignificant for neutrino reactions,
whereas the fourth does not enter in that case and is known to be a minor correction
in the QE regime for electron scattering.

Results obtained within RPWIA show that all scaling functions have
the same shape (see Fig.~\ref{fig:normalized_rpwia}). This comment
also applies to models based on nonrelativistic and semirelativistic descriptions
(see \cite{Amaro05b,Amaro07}).

Within the RMF model, all transverse scaling functions approximately
collapse in a single one. On the contrary, the longitudinal
responses are grouped in two sets: one corresponding to the pure
electron isovector and neutrino (antineutrino) VV-responses, {\it i.e.,}
$f_L^{T=1,ee'}$ and $f_L^{VV,\nu(\bar\nu)}$, and the other to the
isoscalar contribution for electrons, namely, $f_L^{T=0,ee'}$. This
result emerges for all $q$-values and tends to be rather general. It
is also noticeable that the tail is higher and more extended for the
transverse responses, whereas for the longitudinal ones it tends to
go down faster.

It is worth observing that in all cases the RMF scaling functions
display a much more pronounced asymmetric shape than the RPWIA ones,
an effect related to the specific treatment of final state
interactions.

\subsection{Height and position of the peak of the scaling function}\label{shift}

\begin{figure}[htbp]
  \centering
         \includegraphics[width=.34\textwidth,angle=270]{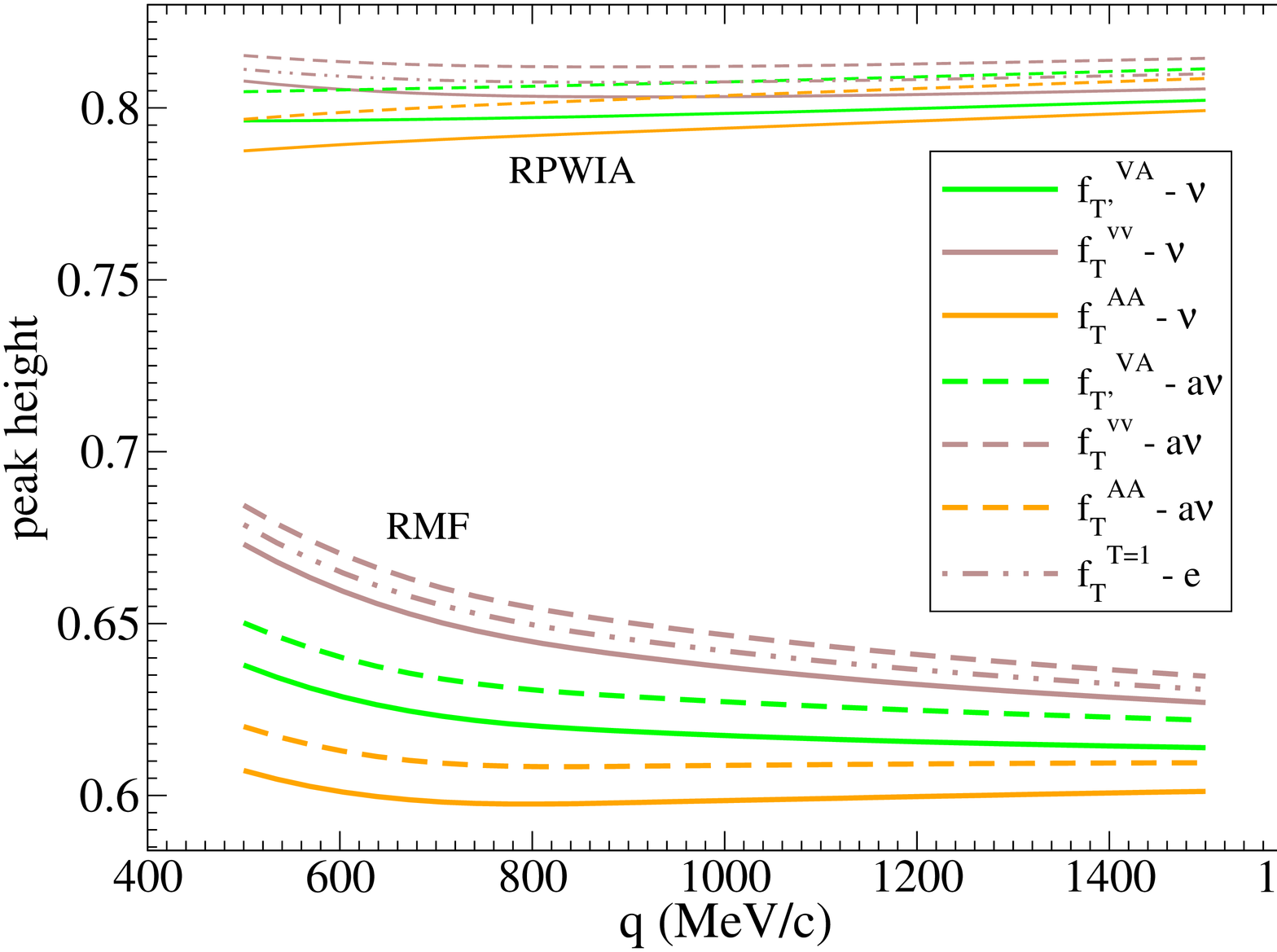}\\
         \includegraphics[width=.34\textwidth,angle=270]{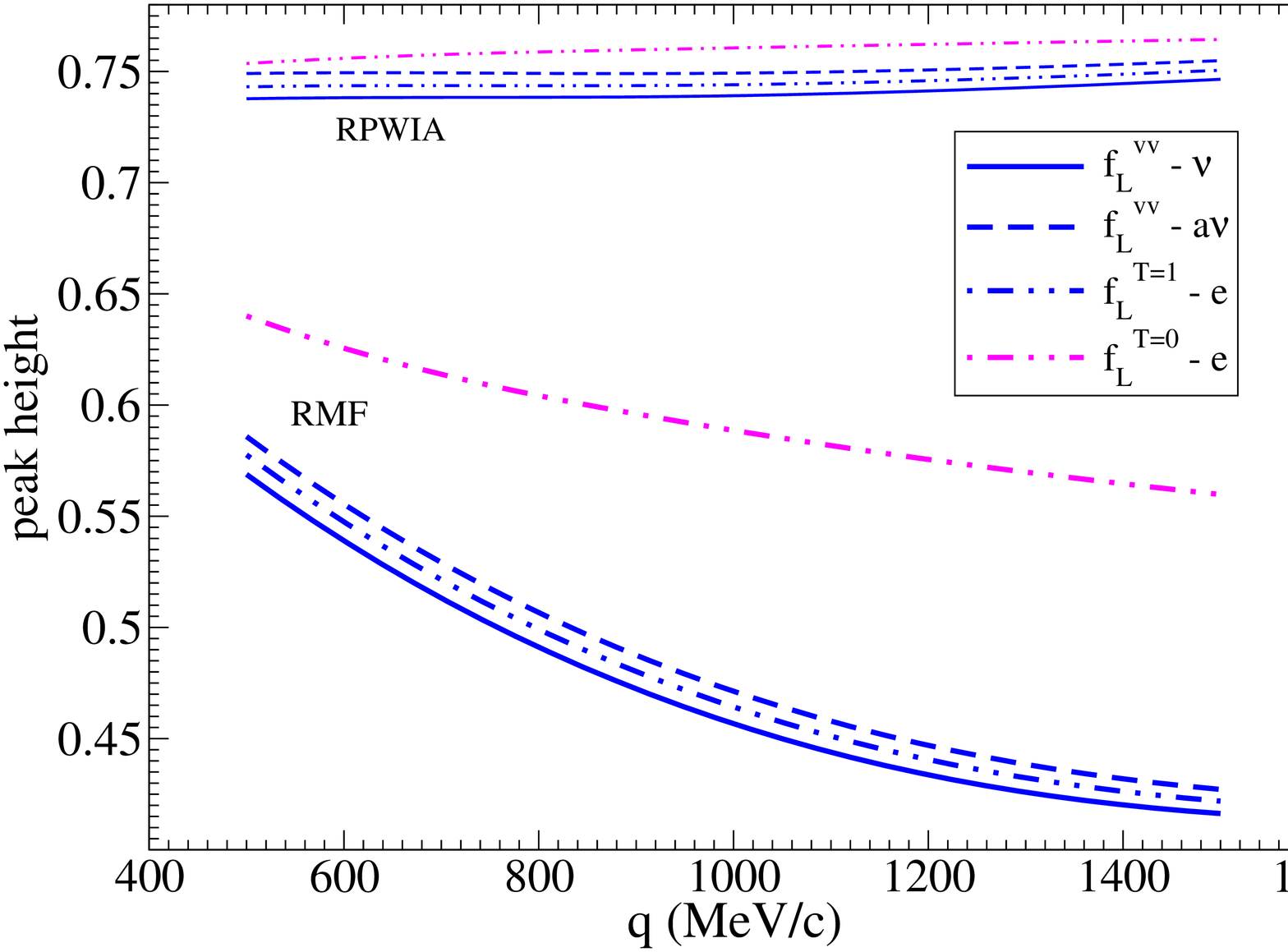}
     \caption{(Top panel) Peak height of the transverse set of scaling functions as
     a function of the transferred momentum $q$.
     The upper set of lines corresponds to the prediction within RPWIA (thin lines), while the lower set of lines has been obtained with the RMF model.
     (Bottom panel) As for the top panel, but now for the longitudinal set of scaling functions.}
     \label{fig:PH}
\end{figure}

In the top (bottom) panel in Fig.~\ref{fig:PH} the peak-height of the transverse
(longitudinal) set of scaling functions is presented as function of $q$.
The results correspond to RMF and RPWIA predictions.
We observe that the peak-heights of the scaling functions within
RPWIA are almost $q$-independent (and very close to RFG value of
3/4), while the RMF ones present a mild $q$-dependence in the
transverse set and a somewhat stronger one for the longitudinal set.
It is well known that FSI tend to decrease the peak-height of the
responses putting the strength in the tails, especially at high energy loss. This is particularly
true for the RMF approach~\cite{Maieron03,Caballero06} and models
based on the Relativistic Green Function
(RGF)~\cite{Meucci03,Meucci09}. Similar effects have also
been observed within semirelativistic approaches~\cite{Amaro05b,Amaro07}.
More specifically, in Fig.~\ref{fig:PH}, we see that the
discrepancies between the RMF and RPWIA peak-height results average
to $\sim$25$\%$ in the transverse set. On the other hand, those
discrepancies are more strongly $q$-dependent in the longitudinal
sector, reaching $\sim$30$\%$ ($\sim$70$\%$) in the lower (higher)
$q$-region for the longitudinal isovector responses (blue lines).
Finally, the difference between the isoscalar longitudinal $(e,e')$
scaling function produced by RMF and RPWIA (magenta dashed-dotted
lines) is somewhat smaller: $\sim$20$\%$ ($\sim$30$\%$) for lower
(higher) $q$.

\begin{figure}[htbp]
  \centering
         \includegraphics[width=.34\textwidth,angle=270]{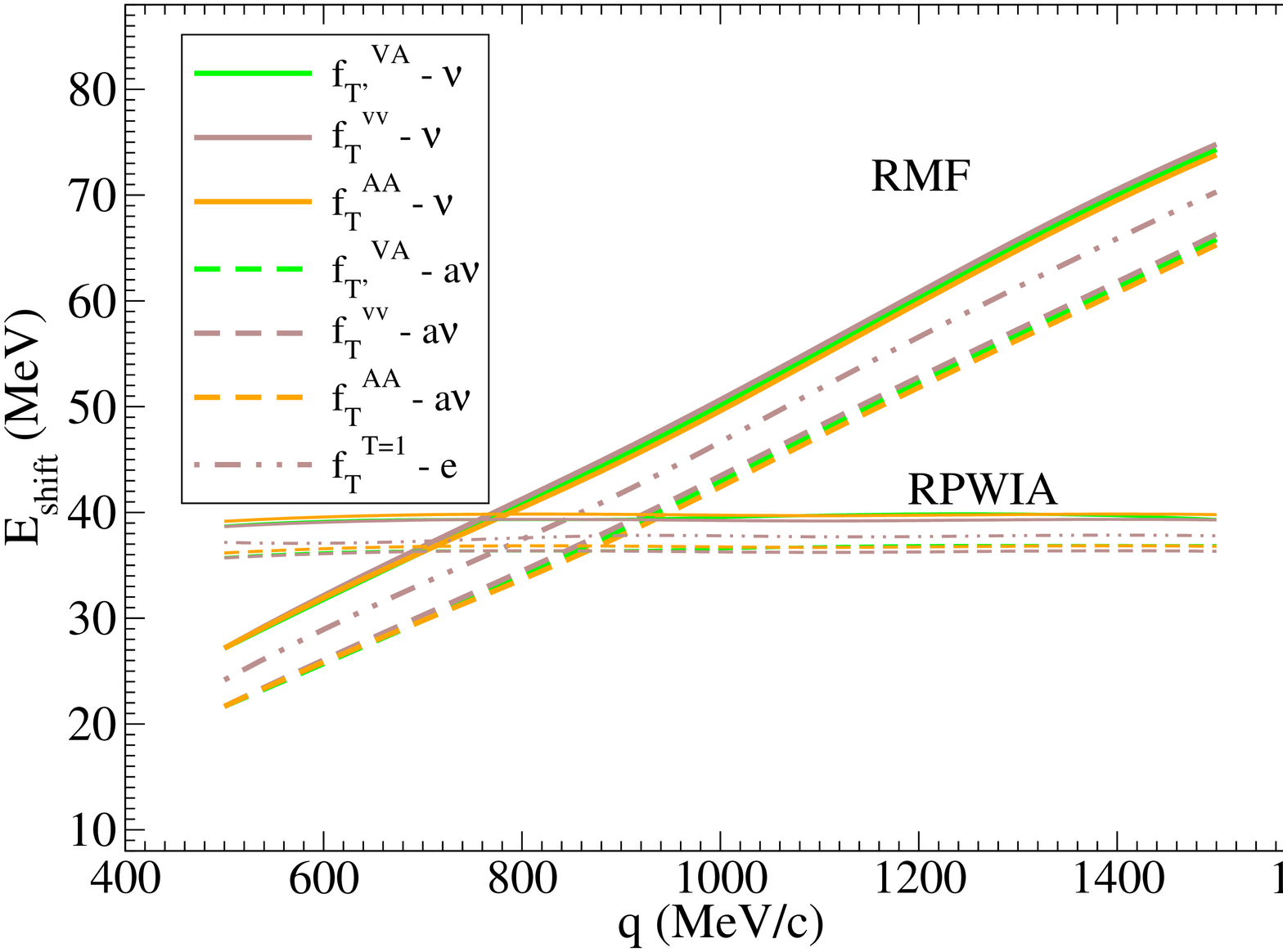}\\
         \includegraphics[width=.34\textwidth,angle=270]{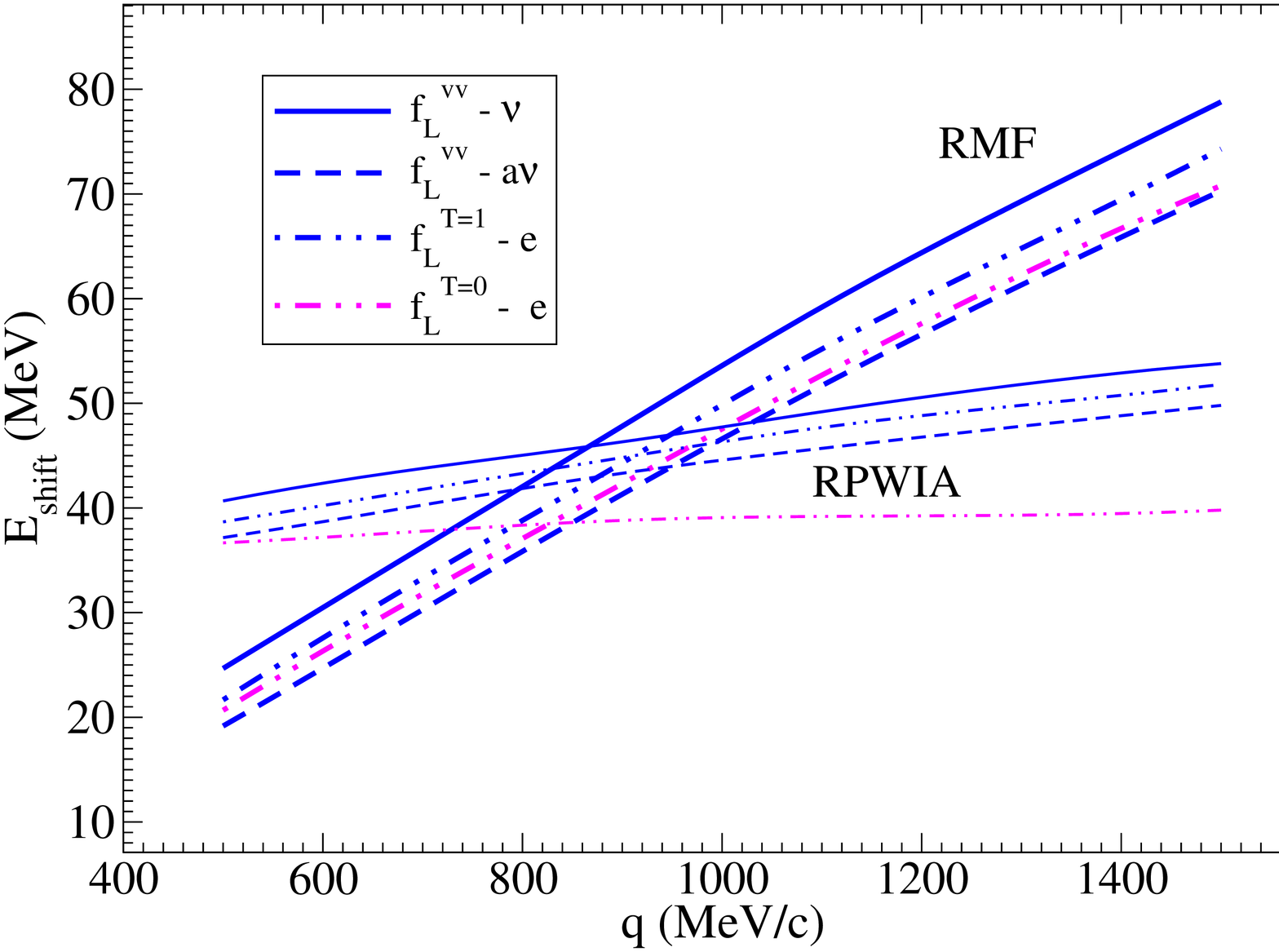}
     \caption{(Top panel) Shift energy, $E_{shift}$, needed in order to have the corresponding scaling function peak located at $\psi'=0$, as function of $q$. Results for the transverse set of scaling functions.
     (Bottom panel) As for the top panel, but now for the longitudinal set of scaling functions.}
     \label{fig:Esh}
\end{figure}

In Fig.~\ref{fig:Esh} we study the position of the peak of the
transverse and longitudinal sets. To this scope we display the
energy shift, $E_{shift}$, needed to place the peak of the scaling
function at $\psi'=0$ as a function of $q$.
In the top panel of Fig.~\ref{fig:Esh} we see that for the RPWIA
transverse scaling function, $E_{shift}$ is almost $q$-independent,
while the corresponding RMF shift increases almost linearly with the
momentum transfer. This $q$-linear dependence of $E_{shift}$ was
already observed and discussed within the framework of a
semirelativistic model based on the use of the
Dirac-equation-based potential~\cite{Amaro07}. Approximately
the same behavior is observed for the longitudinal set (bottom panel
in Fig.~\ref{fig:Esh}), although in this case the RPWIA results are
softly linearly dependent on $q$.
It is also worth mentioning that the three transverse scaling
functions linked to the same neutrino or antineutrino process,
$f_T^{VV}$, $f_T^{AA}$ and $f_{T'}^{VA}$, collapse in a single line
for RMF as for RPWIA.

From the analysis of Figs.~\ref{fig:PH} and \ref{fig:Esh} one may
conclude that $f_L^{T=1,ee'}$ presents the same behavior (height and
position) as $f_L^{VV,\nu(\bar{\nu})}$ (blue lines). The differences
between these three curves are approximately constant and arise from
the differences in the bound states involved in the reaction:
proton+neutron in $(e,e')$, neutron in $(\nu,\mu^-)$ and proton in
$(\bar{\nu},\mu^+)$. The Coulomb-FSI, namely, the electromagnetic
interaction between the struck nucleon and the residual nucleus,
which plays a role when the outgoing nucleon is a proton, could also
introduce a difference; however, we find that its effects are
negligible and that the differences between, for instance, $f_L^{VV,\nu}$ and
$f_L^{VV,\bar{\nu}}$ in RPWIA (where no Coulomb-FSI are involved)
are almost the same as in RMF (see Figs.~\ref{fig:PH} and
\ref{fig:Esh}).

As mentioned in the Introduction, the strong $q$-dependence of the
RMF peak position, which keeps growing with the momentum transfer,
is a shortcoming of the model, whose validity is
questionable at very high $q$. Indeed for high $q$ the outgoing
nucleon carries a large kinetic energy so the effects of FSI should
be suppressed for such kinematics. In fact, it would be
desirable that the RMF results tend to approach the RPWIA ones for
increasing momentum transfer, {\it i.e.,} the scaling functions
should become more symmetric, and a saturation of the peak-height
reduction and of the energy shift should be observed. That trend is
consistent with the scaling
arguments~\cite{Donnelly99a,Maieron02,Caballero06},
{\it i.e.}, the 
experimental evidence
of a universal scaling function for increasing $q$.
This is one of the motivations to use an alternative model if one
aims to reproduce the experimental $(e,e')$ data at medium-to-high momentum
transfers.

A possible alternative for the behavior of the peak height, peak
position and shape of the scaling functions would be to implement
the RMF model at low to intermediate-$q$ and the RPWIA one for higher
$q$-values.


\subsection{Sum rules}\label{integrals}

\begin{figure}[htbp]
   \centering
         \includegraphics[width=0.35\textwidth,angle=270]{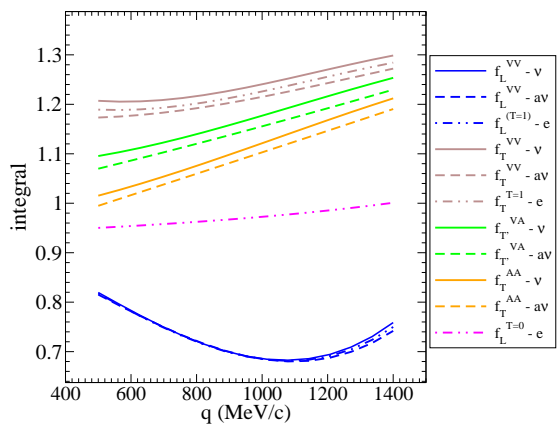}
     \caption{Integrals of RMF scaling functions as functions of $q$.}
     \label{fig:integrals_rmf}
\end{figure}

In Fig.~\ref{fig:integrals_rmf}, the values of the integrals over
$\psi'$ of the different scaling functions within RMF model are
presented versus $q$. These are given by
\begin{equation}
S_i(q) = \int_{-\infty}^{\infty} f_i(\psi,q) \,d\psi \,.
\end{equation}
The integration limits, denoted by $(-\infty,+\infty)$, extend in
reality to the range allowed by the kinematics. The above integral
in the case of the longitudinal $(e,e')$ scaling function was shown
to coincide, apart from some minor discrepancies ascribed to the
particular single-nucleon expressions considered and the influence
of the nuclear scale introduced, with the results obtained using the
standard expression for the Coulomb Sum Rule (see
\cite{Caballero10a} for details). Hence in what follows we denote
the functions $S_{i}(q)$ simply as sum rules.

We see that all integrals of the transverse set are above unity and
increase almost linearly with $q$. On the contrary, the integrals of
$f_L^{VV,\nu(\bar\nu)}$ and $f_L^{T=1,ee'}$ (blue lines) are below
unity and decrease with $q$ up to $q=1100$ MeV/c. From $q=900$ MeV/c
they begin to be stable around the value $0.7$. Then, from $q=1200$
MeV/c to higher $q$-values the integrals start growing again. However,
notice that in that $q$-region the result of the integrals is very
sensitive to the behavior of the tail of these particular scaling
functions (see Fig.~\ref{fig:normalized_L}). Finally, the values
of the integral of the longitudinal isoscalar function,
$f_L^{T=0,ee'}$, is approximately constant and close to unity. The
behavior of the integrals of the two longitudinal scaling functions
for $(e,e')$ is consistent with the analysis of the Coulomb sum rule
for these two models (see \cite{Caballero10a}).

Although not shown here, we have also studied the integrals within RPWIA.
In general, one observes that they are almost $q$-independent in all cases:
$\sim$1 for the longitudinal set and $\sim$1.05 for the transverse set.

\section{Extension of the SuperScaling Approach: the SuSAv2 model}\label{SuSAv2definition}

In this section we build the SuSAv2 model as a combination of the
original SuSA model and some of the physical ingredients contained
in the RMF  and RPWIA models.

On the one hand, as we have shown in the previous sections, the RMF
model has a $q$-dependence that is too strong. On the other hand, the SuSA
model does not account for the difference between the longitudinal
and transverse $(e,e')$ scaling functions. Similarly, SuSA does not
account for possible differences in the scaling function linked to
isospin effects (isovector, isoscalar, isovector+isoscalar) or to
the character of the current ($J_VJ_V$: vector-vector, $J_VJ_A$:
axial-vector, $J_AJ_A$: axial-axial).

Thus, we aim to improve the SuSA model by introducing into it
specific information from the RMF approach. The goal is to get a new
version of SuSA, SuSAv2. The model is based on the following four
assumptions:
\begin{enumerate}
 \item $f_L^{ee'}$ superscales,
 {\it i.e}, it is independent of the momentum
 transfer (scaling of first kind) and of the nuclear species (scaling of second kind).
 It has been proven that $f_L^{ee'}$ superscales for a range of $q$ relatively
 low ($300 < q <570$ MeV/c), see \cite{Donnelly99a}.
 As in the original SuSA model, here we
assume that superscaling is fulfilled by Nature.

 \item $f_T^{ee'}$ superscales.
 It has been shown that $f_T^{ee'}$ approximately superscales in the region 
 $\psi<0$ for a wide range of $q$ ($400<q<4000$ MeV/c), see \cite{Maieron02}.
 However we assume that once the contributions from non-QE processes are removed 
 (MEC, $\Delta$-resonance, DIS, {\it etc.}) the superscaling behaviour could 
 be extended to the whole range of $\psi$.

 \item The RMF model reproduces quite well the relationships between all scaling 
 functions in the {\it whole} range of $q$.
 This assumption is supported by the fact that RMF model is able to reproduce the experimental scaling function, $f_{L,\text{exp}}^{ee'}$, and the fact that it 
 naturally yields the inequality  $f_T^{ee'}>f_L^{ee'}$.

 \item At very high $q$ the effects of FSI disappear and all scaling functions must approach the RPWIA results.

\end{enumerate}

Contrary to what is assumed in the SuSA model, where only
$f_{L,\text{exp}}^{ee'}$ is used as {\it reference} scaling function
to build all nuclear responses, within SuSAv2 we use three RMF-based {\it
reference} scaling functions (which will be indicated with the
symbol $\tilde f$): one for the transverse set, one for the
longitudinal isovector set and another one to describe the
longitudinal isoscalar scaling function in electron scattering. This
is consistent with the study of the shape of the scaling functions
discussed in the previous section, where three different sets of
scaling functions emerged.

\begin{figure}[htbp]
   \centering
         \includegraphics[width=0.33\textwidth,angle=270]{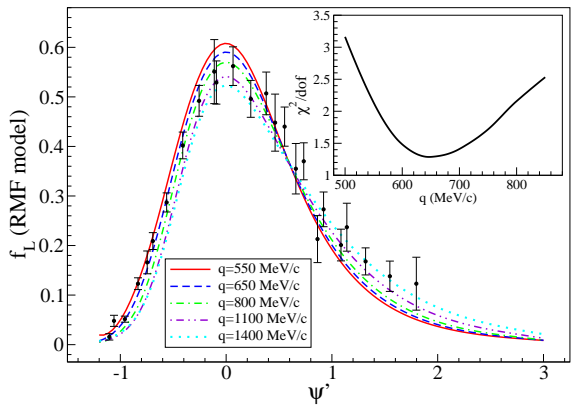}
     \caption{Longitudinal scaling function for $(e,e')$ computed within RMF.
     The scaling functions have been shifted to place the maximum at $\psi'=0$.
     In the inner smaller plot the reduced-$\chi^2$, defined as
     $\chi^2/25 = \frac{1}{25}\sum_{i=1}^{25}[(f^{ee'}_{L,exp,i} - f_{L,i}^{RMF})/\sigma_{L,i}^{exp}]^2$ where $\sigma_{L,i}^{exp}$ are the errors of the experimental data,
     is presented versus $q$. The minimum $\chi^2$ is around $q=650$ MeV/c.
     Data from Ref.~\cite{Jourdan96}.}
     \label{fig:fLRMF}
\end{figure}

We employ the experimental scaling function $f_{L,\text{exp}}^{ee'}$
as guide in our choices for the {\it reference} ones.
In Fig.~\ref{fig:fLRMF} we display the RMF longitudinal scaling
function, $f_L$, for several representative values of $q$. Notice
that the functions have been relocated by introducing an energy shift 
(see later) so that the maximum is at $\psi'=0$. It appears that scaling 
of first kind is not perfect and some $q$-dependence is observed. 
Although all the curves are roughly
compatible with the experimental error bars, the scaling function
that produces the best fit to the data corresponds to $q\approx 650$
MeV/c. This is the result of a $\chi^2$-fit to the 25 experimental
data of $f^{ee'}_{L,exp}$, as illustrated in the inner plot in
Fig.~\ref{fig:fLRMF}.

According to this result, we identify the reference scaling functions with $f^{T=1,ee'}_L$, $f^{T=0,ee'}_L$ and $f^{T=1,ee'}_T$ evaluated within the RMF model at $q=650$ MeV/c and relocated so that the maximum is at $\psi'=0$ (we will account for the energy shift later):
\begin{eqnarray}
 \tilde f_T &\equiv& f_T^{T=1,ee'}|^{RMF}_{q=650} \\
 \tilde f_{L,T=1} &\equiv& f_L^{T=1,ee'}|^{RMF}_{q=650} \\
 \tilde f_{L,T=0} &\equiv& f_L^{T=0,ee'}|^{RMF}_{q=650}\,.
\end{eqnarray}
Thus, by construction, the $(e,e')$ longitudinal scaling function built within SuSAv2 is
$f_L|^{SuSAv2}=f_L|^{RMF}_{q=650}\approx f_{L,exp}^{ee'}$.
In order to work with these reference scaling functions we need analytical expressions for them.
To that end, we have used a skewed-Gumbel function which depends on four parameters.
The expressions that parametrize the reference scaling functions are presented in Appendix~\ref{appendixSG}.


The next step before building the responses (see Eqs.~(\ref{RLTee}-\ref{RTp})) is to define the rest of scaling
functions starting from the reference ones.
According to the third assumption for the construction of SuSAv2,
we define:
\ba
\label{ratio1}
f_L^{VV,\nu(\bar{\nu})}(q) = \mu_L^{VV,\nu(\bar{\nu})}(q) \tilde f_{L,T=1} \\
\label{ratio2}
f_T^{VV,\nu(\bar{\nu})}(q) = \mu_T^{VV,\nu(\bar{\nu})}(q) \tilde f_{T} \\
\label{ratio3}
f_T^{AA,\nu(\bar{\nu})}(q) = \mu_T^{AA,\nu(\bar{\nu})}(q) \tilde f_{T} \\
\label{ratio4}
f_{T'}^{VA,\nu(\bar{\nu})}(q) = \mu_T^{VA,\nu(\bar{\nu})}(q) \tilde f_{T}\,,
\ea
where we have introduced the ratios $\mu$ defined as:
\ba
 \mu_T^{VV,\nu(\bar{\nu})}(q)
      &\equiv&f_T^{VV,\nu(\bar{\nu})}(q) / f_T^{T=1,ee'}(q)  \\
 \mu_T^{AA,\nu(\bar{\nu})}(q)
      &\equiv&f_T^{AA,\nu(\bar{\nu})}(q) / f_T^{T=1,ee'}(q)  \\
 \mu_{T'}^{VA,\nu(\bar{\nu})}(q)
      &\equiv&f_{T'}^{\nu(\bar{\nu})}(q) / f_T^{T=1,ee'}(q) \,,
\ea
for the transverse set and
\ba
  \mu_L^{VV,\nu(\bar{\nu})}(q)
      \equiv f_L^{VV,\nu(\bar{\nu})}(q) / f_L^{T=1,ee'}(q) \,,
\ea
for the longitudinal one.

  \begin{figure}[htbp]
         \includegraphics[width=0.33\textwidth,angle=270]{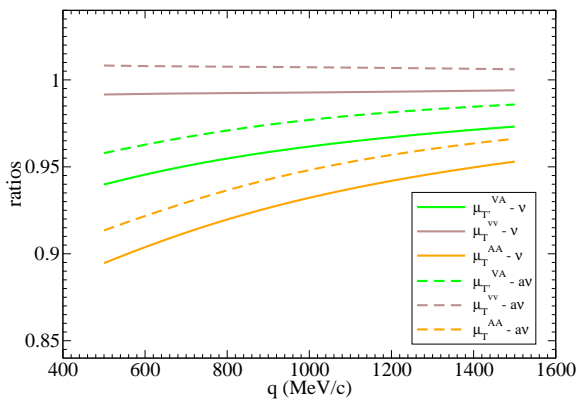}
     \caption{Ratios of transverse scaling functions.}
     \label{fig:ratios_T}
  \end{figure}

From the results of these ratios, presented in
Fig.~\ref{fig:ratios_T}, it emerges that one can assume
$\mu_T^{VV,\nu(\bar{\nu})}(q)\approx1$, with an error of the order
of $\sim$1$\%$. The same assumption could be made for
$\mu_{T'}^{\nu(\bar{\nu})}(q)$ and $\mu_T^{AA,\nu(\bar{\nu})}(q)$
but in this case the error averages to $\sim$3$\%$ and $\sim$7$\%$,
respectively. Regarding the longitudinal isovector set, although not
shown, one gets $\mu_L^{VV,\nu(\bar{\nu})}\approx1$ with an error of
the order $\sim$1$\%$.

Therefore it is a good approximation to set all of the $\mu$-ratios
equal to unity in
Eqs.~(\ref{ratio1},\ref{ratio2},\ref{ratio3},\ref{ratio4}).


In summary, within SuSAv2 we will assume: $f_T^{VV,\nu(\bar{\nu})} =
f_T^{AA,\nu(\bar{\nu})} = f_{T'}^{VA,\nu(\bar{\nu})} = \tilde f_T$
and $f_L^{VV,\nu(\bar{\nu})}=\tilde f_L$.
Notice that since $f_T^{T=0,ee'}$ and
$f_{CC,CL,LL}^{AA,\nu(\bar{\nu})}$ are not defined (see
Sect.~\ref{shape}) we will also assume $f_T^{T=0,ee'}= \tilde
f_{L,T=1}$ and $f_{CC,CL,LL}^{AA,\nu(\bar{\nu})}= \tilde f_{L,T=1}$.

Finally, in order to implement the assumption number 4 of the model,
namely the disappearance of FSI at high $q$, we build the SuSAv2 L
and T scaling functions as linear combinations of the RMF-based and
RPWIA reference scaling functions:
\begin{eqnarray}
{\cal F}_L^{T=0,1} &\equiv& \cos^2 \chi(q) \tilde f_L^{T=0,1} +
\sin^2 \chi(q) \tilde f_L^{RPWIA}  \nonumber\\
{\cal F}_T &\equiv& \cos^2 \chi(q) \tilde f_T +
\sin^2 \chi(q) \tilde f_T^{RPWIA} \,, \nonumber\\
\end{eqnarray}
where
$\chi(q)$ is a $q$-dependent angle given by
\begin{equation}
\chi(q) \equiv \frac{\pi}{2} ( 1-[1+\exp{\left((q-q_0)/w_0\right)}]^{-1} )
\end{equation}
with $q_0$=800 MeV/c and $w_0$=200 MeV.
The reference RPWIA scaling functions, $\tilde f_K^{RPWIA}$, are
evaluated at $q$=1100 MeV/c, while the reference RMF scaling
functions, $\tilde f_K$, are evaluated at $q$=650 MeV/c (see
discussion in Sect.~II). The explicit parametrization of $\tilde
f_K^{RPWIA}$ is given in Appendix B.
With this procedure we get a description of the responses based on
RMF behavior at lower-$q$ while for higher momentum transfers it
mimics the RPWIA trend.
The transition between RMF and RPWIA behaviors occurs at intermediate 
$q$-values, namely, $\sim q_0$, in a region of width $\sim w_0$.

The response functions (see Eqs.~(\ref{RLTee}) and
(\ref{RLVV}--\ref{RTp})) are simply built as: \ba
 R_L^{ee'}(q,\omega) &=& \frac{1}{k_F}
    \left[ {\cal F}_{L,T=1}(\psi')  G_L^{T=1}(q,\omega)\right.\nonumber\\
    &+& \left.  {\cal F}_{L,T=0}(\psi') G_L^{T=0}(q,\omega)\right]
             \label{RLee}\\
 R_T^{ee'}(q,\omega) &=& \frac{1}{k_F}
       {\cal F}_T(\psi') \left[G_T^{T=1}(q,\omega)\right.\nonumber\\
    &+& \left. G_T^{T=0}(q,\omega)\right]
      \label{RTee}
\ea
\ba
 R_L^{VV,\nu(\bar{\nu})}(q,\omega) = \frac{1}{k_F}
   {\cal F}_{L,T=1}(\psi') G_L^{VV}(q,\omega)\\
 R_{CC}^{AA,\nu(\bar{\nu})}(q,\omega)= \frac{1}{k_F}
    {\cal F}_{L,T=1}(\psi') G_{CC}^{AA}(q,\omega) \\
 R_{CL}^{AA,\nu(\bar{\nu})}(q,\omega) = \frac{1}{k_F}
    {\cal F}_{L,T=1}(\psi') G_{CL}^{AA}(q,\omega) \\
 R_{LL}^{AA,\nu(\bar{\nu})}(q,\omega) = \frac{1}{k_F}
    {\cal F}_{L,T=1}(\psi') G_{LL}^{AA}(q,\omega)
%
\ea
\ba
 R_T^{\nu(\bar{\nu})}(q,\omega) &=& \frac{1}{k_F}
             {\cal F}_T(\psi') \left[G_T^{VV}(q,\omega)\right.\nonumber\\
    &+& \left. G_T^{AA}(q,\omega)\right]
             \label{RTnu}\\
 R_{T'}^{\nu(\bar{\nu})}(q,\omega) &=& \frac{1}{k_F}
     {\cal F}_T(\psi')G_{T'}^{VA}(q,\omega).
    \label{RT'nu}
\ea

Furthermore, in order to reproduce the peak position of RMF and
RPWIA scaling functions, discussed in Sect.~\ref{shift}, within
SuSAv2 we consider a $q$-dependent energy shift, namely,
$E_{shift}(q)$. This quantity modifies the scaling variable
$\psi(q,\omega)\longrightarrow\psi'(q,\omega,E_{shift})$ as
described in Appendix~\ref{apendiceScaling}. In particular, we build
this function $E_{shift}(q)$ from the results of the RMF and RPWIA
models presented in Fig.~\ref{fig:Esh}. Thus, $E_{shift}(q)$ for the
reference RMF scaling function $\tilde f_T[\psi'(E_{shift})]$ is the
parametrization of the brown dot-dot-dashed line in the top panel of
Fig.~\ref{fig:Esh}. The same procedure is used to parametrize
$E_{shift}(q)$ corresponding to the $\tilde f_{L,T=1}$ and $\tilde
f_{L,T=0}$, but in this case using, as an average, the blue
dot-dot-dashed line from the bottom panel of Fig.~\ref{fig:Esh}.
Moreover, for the RPWIA case we use for the longitudinal and 
transverse responses the corresponding RPWIA $E_{shift}(q)$ curves 
shown in Fig.~\ref{fig:Esh}.

Notice that for $q\lesssim300-350$ MeV/c it is
difficult to extract the peak position of the RMF scaling function
from the data so we have set a minimum shift energy, $E_{shift}=10$ MeV.
This choice of $E_{shift}(q)$ depending on the particular $q$-domain
region considered is solely based on the behavior of the
experimental cross sections and their comparison with our
theoretical predictions (see results in next sections). In the past
we have considered a fixed value of $E_{shift}$ (different for each
nucleus) to be included within the SuSA model in order to fit the
position of the QE peak for some specific $q$-intermediate values.
Here we extend the analysis to very different kinematics covering
from low to much higher $q$-values. On the other hand, the RMF model
leads the cross section to be shifted to higher values of the
transferred energy. This shift becomes increasing larger for higher
$q$-values as a consequence of the strong, energy-independent,
highly repulsive potentials involved in the RMF model. Comparison
with data (see the results in the next sections) shows that the
shift produced by RMF is too large. Moreover, at very high
$q$-values, one expects FSI effects to be less important and lead to
results that are more similar to those obtained within the RPWIA
approach. This is the case when FSI are described through
energy-dependent optical potentials. Therefore, as already
mentioned, our choice for the functional dependence of
$E_{shift}(q)$ is motivated as a compromise between the predictions
of our models and the comparisons with data.

\section{Comparison with electron scattering data}\label{ee-results}

In this section we present a systematic comparison of {\it total}
inclusive $^{12}$C$(e,e')$ experimental cross sections and the
predictions for the QE process within RMF,
SuSA and SuSAv2 models. As mentioned, data correspond to the total 
inclusive cross section which includes contributions from several channels, 
mainly:
QE scattering, inelastic scattering, many-nucleon emission, {\it etc.}
The models presented in this work aim to describe only the QE
process. Therefore, one expects that the models do not reproduce the
total inclusive experimental data corresponding to kinematical
situations in which non-QE contributions play some role.
Thus, the main interest of the systematic analysis presented in this
section is the comparison between SuSAv2 predictions and those from
the SuSA and RMF models.
Full analyses of the inclusive $(e,e')$ cross section (including
descriptions of QE and non-QE contributions) have been presented with
some success in the past~\cite{Barbaro04,Amaro05a,Maieron09}.
We plan to complete the description of the inclusive process within the
context of SuSAv2 model, as was made in
\cite{Amaro05a,Maieron09} within SuSA, in a near future.

In Figs.~\ref{fig:lowerq_r}-\ref{fig:higherq_r} we present the
comparison of the $(e,e')$ experimental data and models. Due to the
large amount of available data on $^{12}$C$(e,e')$ at different
kinematics (see \cite{ee-data,Benhar:2006wy}) in these three
figures we only show some representative examples.
Each figure is labeled by the incident electron energy,
$\varepsilon_i$ (in MeV), the scattering angle, $\theta_e$, and the
transferred momentum corresponding to the center of the quasielastic
peak, $q$ (in MeV/c).
Pauli Blocking has been included in the SuSA and SuSAv2 models
following the procedure described in
\cite{Rosenfelder:1980nd,Megias14}. In
Appendix~\ref{appendixPB} we present a comparison of the models
(SuSA and SuSAv2) and data when PB is or is not included.
The panels in Figs.~\ref{fig:lowerq_r}-\ref{fig:higherq_r} are
organized according to the value of the transferred momentum (at the
center of the QE peak) in three sets: low-$q$ (from $q=238$ to
$q=333$ MeV/c) in Fig.~\ref{fig:lowerq_r}, medium-$q$ (from $q=401$ to
$q=792$ MeV/c) in Fig.~\ref{fig:mediumq_r} and, high-$q$ (from $q=917$
to $q=3457$ MeV/c) in Fig.~\ref{fig:higherq_r}.
The only phenomenological parameters entering in the calculation are
the Fermi momentum $k_F$ and the energy shift $E_{shift}$. For these
we use $k_F=228$ MeV/c (see \cite{Maieron02}) in both SuSA and
SuSAv2 models. A constant energy shift of 20 MeV is employed in
SuSA~\cite{Maieron02} while a $q$-dependent function, the one
described in Sect.~\ref{SuSAv2definition}, is used for $E_{shift}$
in the SuSAv2 model.

  \begin{figure}[htbp]
         \includegraphics[width=.48\textwidth,angle=0]{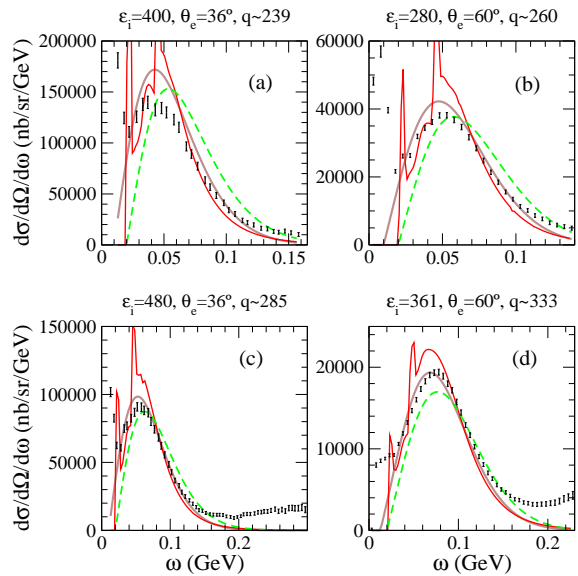}
     \caption{Comparison of inclusive $^{12}$C$(e,e')$ cross sections and predictions of the RMF (red), SuSA (green-dashed) and SuSAv2 (brown) models (see text for details).
     Set of panels corresponding to low-$q$ values.
     Data taken from \cite{ee-data}.}
     \label{fig:lowerq_r}
  \end{figure}

  \begin{figure}[htbp]
         \includegraphics[width=.48\textwidth,angle=0]{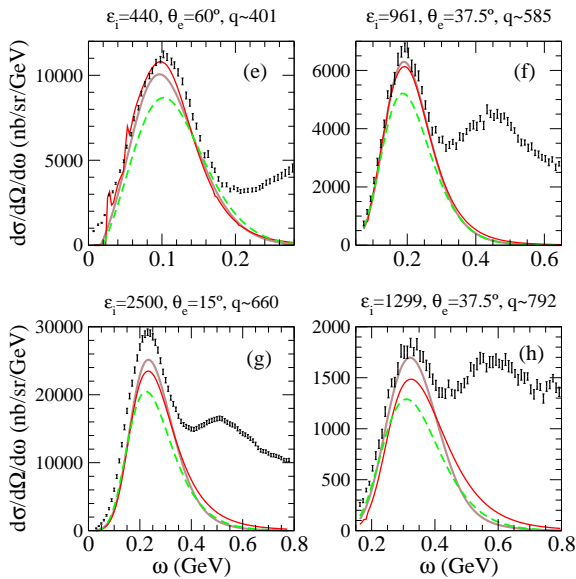}
     \caption{Continuation of Fig.~\ref{fig:lowerq_r}.
     Set of panels corresponding to medium-$q$ values. }
     \label{fig:mediumq_r}
  \end{figure}

  \begin{figure}[htbp]
         \includegraphics[width=.48\textwidth,angle=0]{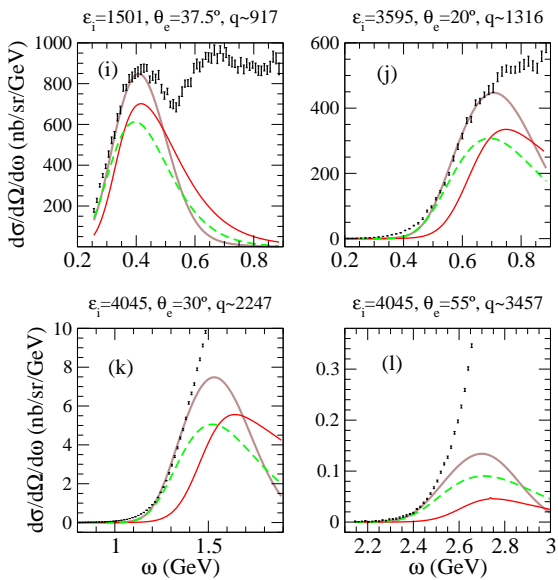}
     \caption{Continuation of Fig.~\ref{fig:lowerq_r}.
     Set of panels corresponding to high-$q$ values. }
     \label{fig:higherq_r}
  \end{figure}


We begin commenting on the low-$q$ panels presented in
Fig.~\ref{fig:lowerq_r}. The main contributions to the cross section
from non-QE processes such as inelastic processes contributions
($\Delta$-resonance) and MEC, are very small, even negligible, in
this low-$q$ region. In spite of that, when the transferred energy
is small ($\omega\lesssim50-60$ MeV) other processes such as
collective effects contribute to the cross section making
questionable the treatment of the scattering process in terms of
IA-based models. This could explain, in
part, the general disagreement between models and data in that
$\omega$ region in (a), (b) and (c) panels.

Some clarifications are called for regarding the RMF results in
Fig.~\ref{fig:lowerq_r}, where sharp resonances appear  at very low
$\omega$ values. These correspond to 1p1h excitations with the phase
shift of a given partial wave going through 90 degrees. With more
complicated many-body descriptions these sharp features are smeared
out.

In summary, in order to test the goodness of the models in the
kinematical situation of Fig.~\ref{fig:lowerq_r}, one should focus
on the study of the tails of the cross sections where large enough
$\omega$-values ($\omega\gtrsim50-60$ MeV) are involved. There, one
observes that SuSA predictions are clearly over-shifted to high
$\omega$-values while RMF and SuSAv2 models fit the
data reasonably well. In addition, as expected, SuSA results are systematically
below SuSAv2 and RMF ones at the QEP.


We now discuss the results for medium-$q$ values presented in
Fig.~\ref{fig:mediumq_r}. First of all, one should mention that for
the kinematics of this figure, in addition to the QE process, non-QE
contributions are essential to describe the experimental cross
sections. For instance, in panels (f), (g) and (h)  the
$\Delta$-peak appears clearly defined at $\omega$ values above the
QE peak.
In panel (e) one sees that
 in the region around the center of the QE-peak,
the RMF prediction is above the SuSAv2 one, being closer to the
experimental data. This is consistent with the behavior of the RMF
scaling function studied in Sect.~\ref{scalingRMF} (see
Fig.~\ref{fig:PH}), namely, the peak-height of the RMF scaling
functions increases for decreasing $q$-values.

If the main non-QE contributions are not included in the
modeling it is hard to conclude which model is better to
reproduce the purely QE cross section. However, it seems reasonable
to conclude that SuSAv2 improves the agreement with data compared to
SuSA. For instance, in the situation of panel (e), it would be
needed that non-QE processes would contribute more than 20\% to the
total cross section in order to SuSA fits the height of the data
around the center of the QE-peak. A 20\% fraction of the cross section linked
to $\Delta$-resonance and MEC contributions is probably too much for
that kinematics. Similar comments and conclusions apply to the
results in panel (d) of Fig.~\ref{fig:lowerq_r}.

For $q$-values close to $650$ MeV/c (panels (f) and (g)) RMF and
SuSAv2 produce very similar results because of the way in which
SuSAv2 has been defined (see Sect.~\ref{SuSAv2definition}). 
For higher $q$-values, $q\gtrsim792$ MeV/c ((h) panel), SuSAv2 and RMF
predictions begin to depart from each other. 
In particular, RMF results tend to shift the peak to higher $\omega$ values 
and to place more strength in the tail 
while SuSAv2 cross sections tend to be more symmetrical due to 
the increasing dominance of the RPWIA scaling behavior (see
Sect.~\ref{SuSAv2definition}).

This difference is more evident for
higher $q$-values, as observed in panels (j)-(l) of
Fig.~\ref{fig:higherq_r}. It is important to point out that for the
kinematics presented in Fig.~\ref{fig:higherq_r} the non-QE
contributions are not only important but they become dominant in the
cross sections. This is the case presented in panels (k) and (l)
where the QE-peak is not even visible in the data.

We could summarize the main conclusions from the present comparison of 
models and data as follows:
\begin{itemize}
 \item Regarding the enhancement of the transverse response, $R_T$, 
 in SuSAv2 compared with SuSA:
 in the absence of modeling of non-QE contributions, the most clear 
 indications that support the SuSAv2 assumptions arise from the
 comparison with data at kinematical situations in which non-QE effects are 
 supposed to be small (panels (e) and (d) in
 Figs.~\ref{fig:lowerq_r} and \ref{fig:mediumq_r}, respectively).
 \item Regarding the energy shift study:
  within the SuSA model we have used a constant energy shift of 20 MeV/c.
  On the one hand, from the comparison with the low-$q$ set of experimental data, 
  Fig.~\ref{fig:lowerq_r}, one concludes that
  20 MeV is a too large shift.
  On the other hand, the comparison with the high-$q$ set of data,
  Fig.~\ref{fig:higherq_r}, suggests that 20 MeV is probably too small.
  Then, one is led to conclude that a constant energy shift is not the best 
  option to reproduce $(e,e')$ data.
  These results support the idea of introducing a $q$-dependent energy shift 
  such as we made in the SuSAv2 model.
  The theoretical justification of this assumption was already discussed in Sect.~\ref{SuSAv2definition}.
\end{itemize}

\section{Comparison with neutrino and antineutrino data}\label{nu-results}

In recent years a significant amount of charge-changing quasielastic
(CCQE) neutrino and antineutrino cross section data have been
presented in the literature. In this section, as in the
previous one for the $(e,e')$ process, we compare the results of
SuSAv2 model with some selected samples from different experiments:
MiniBooNE~\cite{MiniBooNECC10,MiniBooNECC13},
Miner$\nu$a~\cite{MINERVAnu13,MINERVAnub13} and
NOMAD~\cite{NOMAD09}. The SuSA predictions are also presented as
reference.

MiniBooNE has measured CCQE cross sections that are higher than most
predictions based on IA. The excess, at relatively low energy
($\langle E_\nu \rangle \sim0.7$ GeV), observed in MiniBooNE cross
sections has been interpreted as evidence that non-QE processes
may play an important role at that kinematics~\cite{AmaroMEC,
MartiniMEC, NievesMEC}. It is important to point out that in the
experimental context of MiniBooNE, ``quasielastic'' events are
defined as those from processes or channels containing no mesons in
the final state. Thus, in principle, in addition to the purely QE
process, which in this work refers exclusively to processes induced
by one-body currents (IA), meson exchange current
effects (induced by two-body or many-body currents) should also be
taken into account for a proper interpretation of data.

\begin{figure}[htbp]
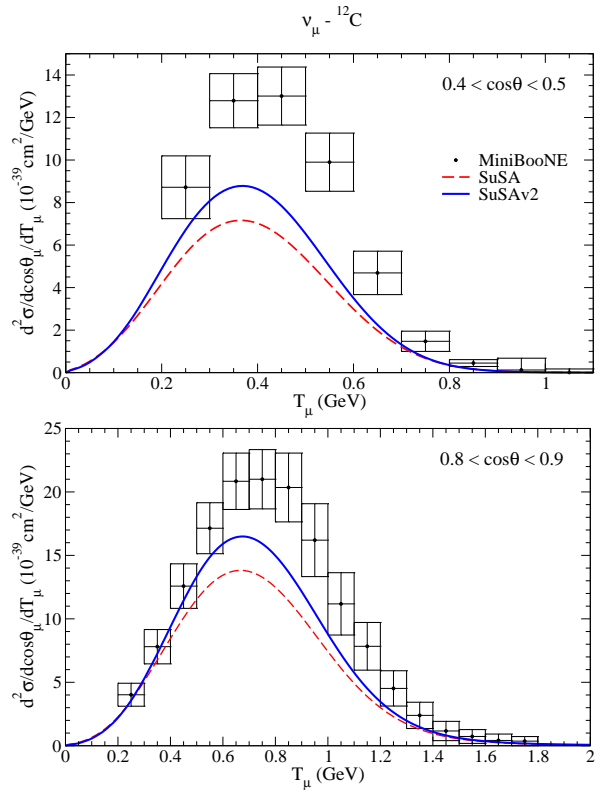

         \includegraphics[width=.48\textwidth,angle=0]{dsdtdc_045cos_band_new.eps}\\
         \includegraphics[width=.48\textwidth,angle=0]{dsdtdc_085cos_band_new.eps}
         \caption{MiniBooNE double differential $(\nu,\mu^-)$ cross section data~\cite{MiniBooNECC10} are compared with SuSA (dashed-red line)
           and SuSAv2 (solid-blue line) predictions.
     In the top panel the scattering angle of the muon is $0.4<\cos\theta<0.5$, while in bottom panel $0.8<\cos\theta<0.9$.}
     \label{fig:MB_nu}
  \end{figure}

\begin{figure}[htbp]
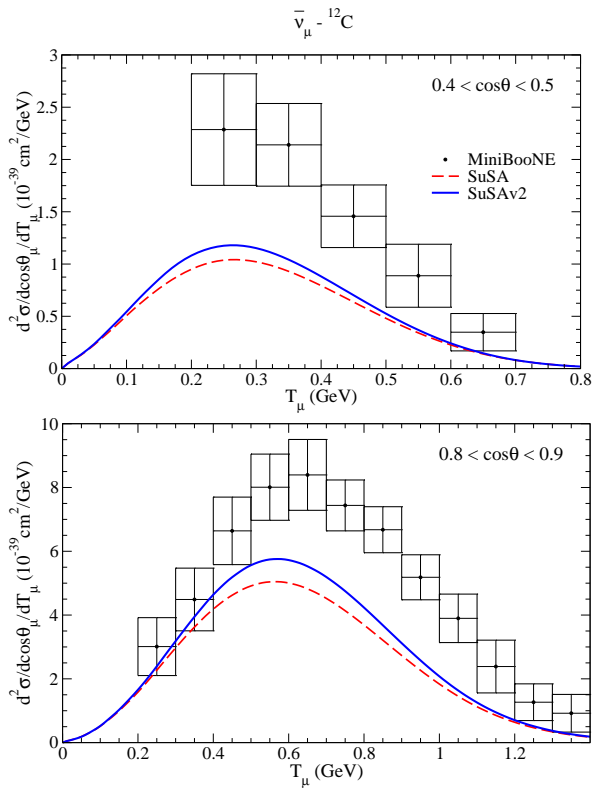

         \includegraphics[width=.48\textwidth,angle=0]{dsdtdc_045cos_bar_band_new.eps}\\
         \includegraphics[width=.48\textwidth,angle=0]{dsdtdc_085cos_bar_band_new.eps}
     \caption{As in Fig.~\ref{fig:MB_nu}, but now for the antineutrino-induced reaction, $(\bar{\nu},\mu^+)$.
     Data taken from \cite{MiniBooNECC13}.}
     \label{fig:MB_nub}
  \end{figure}

In Fig.~\ref{fig:MB_nu} and Fig.~\ref{fig:MB_nub} the double
differential $(\nu,\mu^-)$ and $(\bar{\nu},\mu^+)$ cross sections
measured by the MiniBooNE Collaboration are compared with SuSAv2 (solid-blue
line) and SuSA (dashed-red line) predictions. The top and bottom panels correspond
to a muon scattering angle of $\sim$63$^o$ and $\sim$32$^o$,
respectively. As observed, the SuSAv2 cross section is significantly
larger than SuSA one, although it still falls below the MiniBooNE data.
Thus, there is still room for MEC contributions. In
\cite{Ivanov13} the RMF model is compared with the same set of
data as shown in Figs.~\ref{fig:MB_nu} and \ref{fig:MB_nub}. In
general, one observes that RMF and SuSAv2 models produce almost
identical results (as happened in $(e,e')$ for intermediate
$q$-values).

\begin{figure}[htbp]
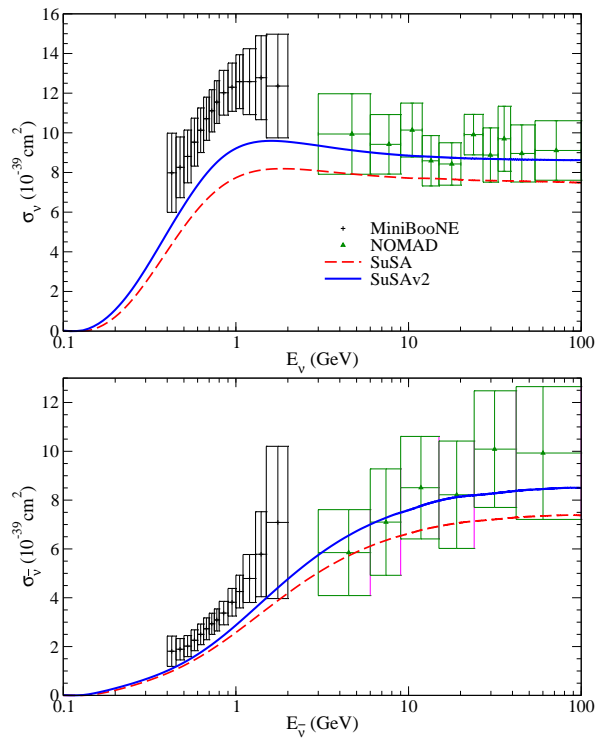

         \includegraphics[width=.48\textwidth,angle=0]{totalnu_band_new.eps}\\
         \includegraphics[width=.48\textwidth,angle=0]{totalnubar_band_new.eps}
     \caption{(Top panel) CCQE $(\nu,\mu^-)$ cross section per nucleon is presented as a function of the incident neutrino energy, $E_\nu$.
     Data from MiniBooNE~\cite{MiniBooNECC10} and NOMAD~\cite{NOMAD09} are compared with SuSA (dashed-red line) and SuSAv2 (solid-blue line) predictions.
     (Bottom panel) As in top panel, but now for the antineutrino-induced reaction, $^{12}$C$(\bar{\nu},\mu^+)$.}
     \label{fig:total_cs}
  \end{figure}

In the NOMAD experiment the incident neutrino (antineutrino) beam
energy is much larger, with a flux extending from $E_\nu$= 3 to 100
GeV. In this case, one finds that data are in reasonable agreement
with predictions from IA models~\cite{Megias13,Ivanov14}. Notice
that the large error bars of these data do not allow for further
definitive conclusions. In Fig.~\ref{fig:total_cs} we present the
CCQE total cross section for neutrino (top panel) and antineutrino
(bottom panel) reactions. Experimental data from NOMAD and MiniBooNE
are compared with SuSA and SuSAv2. SuSAv2 improves the agreement
with the NOMAD data, being, in general, closer to the center of the
bins. The extension of the RMF model to very high energies requires
at first complicated and very long time-consuming calculations. In
this sense, the advantage of SuSAv2 is that it can be easily and
rapidly extended up to very high neutrino energies. Although not
shown here, preliminary results evaluated with the RMF model at
NOMAD kinematics~\cite{UdiasP} are proved to be very similar to the
SuSAv2 ones.

\begin{figure}[htbp]
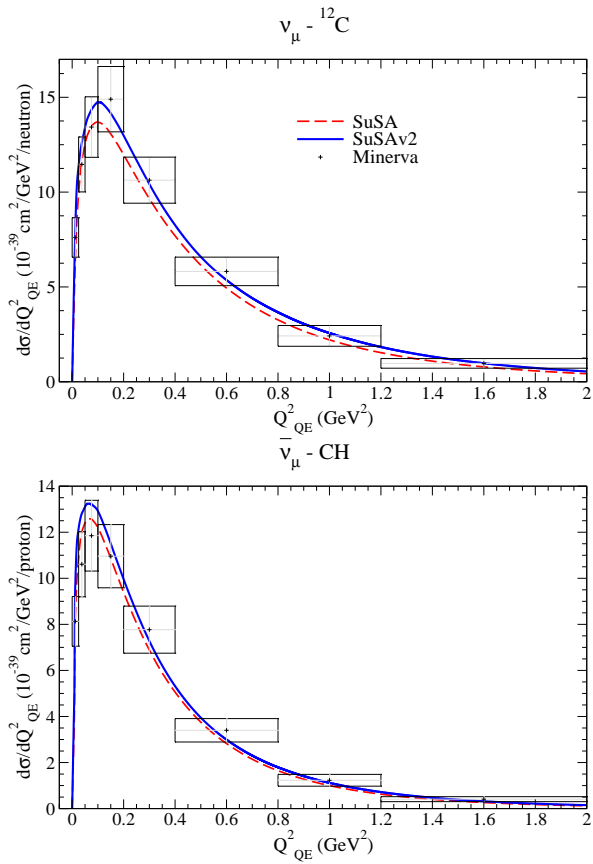

         \includegraphics[width=.48\textwidth,angle=0]{MINERVA_nu.eps}\\
         \includegraphics[width=.48\textwidth,angle=0]{MINERVA_nub.eps}
     \caption{CCQE neutrino (upper panel) and antineutrino (lower panel) MINER$\nu$A data are compare with SuSA (dashed-red line)
       and SuSAv2 (solid-blue line) predictions.
     Data taken from~\cite{MINERVAnu13,MINERVAnub13}.}
     \label{fig:Minerva}
  \end{figure}

In the MINER$\nu$A experiment the neutrino energy flux extends from
1.5 to 10 GeV and is peaked at $E\nu\sim3$ GeV, {\it i.e.,} in between
MiniBooNE and NOMAD energy ranges. Therefore, its analysis can
provide useful information on the role played by meson-exchange
currents in the nuclear dynamics.
In recent work~\cite{Megias14} it was found that, contrary to the
comparison with the MiniBooNE data, the two IA models analyzed (RMF
and SuSA) provide a good description of the MINER$\nu$A data without
the need of significant contributions from MEC. In
Fig.~\ref{fig:Minerva} we present the single-differential cross
section ($d\sigma/dQ^2_{QE}$), measured by MINER$\nu$A, as a function
of the reconstructed four-momentum transfer squared, $Q^2_{QE}$ (see
\cite{MINERVAnu13,MINERVAnub13} for explicit definition of
$Q^2_{QE}$). The SuSA and SuSAv2 results are compared with
MINER$\nu$A data. In spite of the enhancement with respect to SuSA,
SuSAv2 is not only consistent, but it also improves the agreement
with MINER$\nu$A data. In fact, RMF and SuSAv2 models produce very
close results (RMF predictions are presented in
\cite{Megias14}). Thus, contrary to the MiniBooNE situation,
the comparison of MINER$\nu$A data and IA based models, in
particular, RMF and SuSAv2, leaves little room for MEC
contributions.

A further general comment on the previous results is in order: the
difference between SuSA and SuSAv2 is larger for neutrino than for
antineutrino results. This occurs because of the cancellation
occurring between $R_T$ (positive) and $R_{T'}$ (negative) responses
in antineutrino cross sections. Notice that the transverse responses
are substantially enhanced in SuSAv2 compared with SuSA.

In summary we find that SuSAv2 compared with the SuSA model improves
the comparison with neutrino and antineutrino data. Additionally,
SuSAv2 (as SuSA) can easily make predictions at kinematics (very
high energies) in which other more microscopic-based models, as RMF,
require additional assumptions and demanding, time-consuming calculations.

\section{Conclusions}\label{conclusions}

The SuSA model, based on the scaling behavior exhibited by $(e,e')$
data in the longitudinal channel, has been extensively used in the
past not only to explain electron scattering, but also neutrino
reactions. The basic idea of SuSA is the existence of a universal
scaling function, the one ascribed to longitudinal $(e,e')$ data, to
be applied to any other process. Hence, SuSA makes use of the same
scaling function for the two channels, longitudinal and transverse,
involved in QE electron scattering reactions, as well as for the
whole set of responses that enter in charged-current neutrino
scattering processes.

On the other hand, the RMF model provides a description of the
scattering reaction mechanism including the role played by FSI. The
RMF model leads to a longitudinal scaling function in accordance
with data, and hence, also in agreement with the SuSA result.
However, contrary to the main assumption considered by SuSA, namely,
the existence of only one universal scaling function, the RMF model
provides a transverse scaling function that is higher by $\sim$20$\%$
compared with the longitudinal one. In other words, scaling of zeroth
kind is not fulfilled by RMF. This result also seems to be in
accordance with the preliminary analysis of data that shows the pure
QE transverse channel to lead to a scaling function exceeding the
longitudinal one by an amount, $\sim$20-25$\%$, similar to the one
shown by the RMF results.

The analysis of neutrino reactions also introduces basic differences
with the electron case. Whereas in the latter, responses contain
isoscalar and isovector contributions, in the former, the responses
are purely isovector. Moreover, not only pure vector-vector
responses contribute to neutrino processes, but also axial-axial and
the interference axial-vector one. All of these results, in addition to
the preliminary analysis of the separate QE longitudinal and
transverse responses, may introduce some doubts about the existence of
a unique scaling function valid for all processes.

In this work we have pursued this problem and have extended the SuSA
model by taking into account the results provided by the RMF
approach. Hence we study in detail RMF scaling functions
corresponding to all channels, and from this we select the minimum
set of scaling functions, named reference scaling functions, that
allow us to construct the cross section for electron and neutrino
scattering reactions. The new model, called SuSAv2, takes care of
the enhancement of the $(e,e')$ transverse response compared with the
longitudinal one, as well as the general behavior shown by the
functions ascribed to neutrino reactions.

SuSAv2 is based on a ``blend" between the properties of the RMF and
RPWIA responses. The former appears to do well at low to
intermediate values of the momentum transfer, for instance, yielding
both an asymmetric scaling function and the T/L differences observed
in electron scattering data. However, because of the strong energy
independent scalar and vector potentials involved, the RMF model
does less well at high values of $q$, where the energy shift is seen
to be too strong and the high-energy tail in the RMF scaling
function is likely too large. The RPWIA, on the other hand, does not
work well at low to intermediate momentum transfers and, in fact,
yields results that are not very different from those of the
relativistic Fermi gas, which are known to be too symmetrical and
not to contain the T/L differences seen in both the RMF results and
in electron scattering data. What SuSAv2 attempts to do is to
provide a cross-over from the low to intermediate momentum transfer
regime (where the RMF results are employed) to the high-$q$ regime
(where the results revert to those of the RPWIA). A particular,
reasonable ``blending'' function has been used, although the
specific parametrization assumed is not critical. Indeed, when
updated 2p-2h MEC responses and updated representations of inelastic
contributions are incorporated (see below) it will be appropriate to
make detailed fits to existing electron scattering data and at that
point one can refine the determination of the parameters used in
this initial study.

We have applied the new SuSAv2 model to the description of electron
and neutrino scattering, and have proved that SuSAv2 predictions are
higher than the SuSA ones and are closer to data. This is so for
electron scattering as well as for neutrino reactions. However, in the latter,
theory still underestimates data in most of the cases, in
particular, for the kinematics corresponding to the MiniBooNE
experiment. This outcome is similar to the one already observed for
the RMF results.

SuSAv2 model incorporates some basic ingredients not taken into
account within SuSA, hence it clearly improves its reliability to
the description of scattering processes, being at the same time a
model that is easy to implement in the ``generator codes'' used to analyze
the experiments. Moreover, its application to very high energies
does not involve particularly demanding calculations in contrast to the
RMF model that may can complex and long, time-consuming
calculations.

Finally, a comment is in order concerning the ingredients
incorporated by SuSAv2 (likewise for SuSA and RMF). This is a model
based exclusively on the IA. Hence
ingredients beyond the IA, {\it i.e.,} two-body meson-exchange currents,
inelastic contributions, {\it etc.,} should be added to the model. Work
along these lines is presently under way, as well as the application
of SuSAv2 to different experimental kinematics: Argoneut, T2K, {\it etc.}
These studies  will be presented in a forthcoming publication.

\vspace{0.5cm}
This work was partially supported DGI (Spain): FIS2011-28738-C02-01,
by the Junta de Andaluc\'i­a (FQM-170, 225),
by the INFN National Project MANYBODY,
and the Spanish Consolider-Ingenio 2000 programmed CPAN,
in part (T.~W.~D.) by the U.S. Department of Energy under cooperative agreement DE-FC02-94ER40818 and in part (M.~B.~B.) by the INFN under project MANYBODY.
G.~D.~M. acknowledges support from a fellowship from the Fundaci\'on C\'amara (Universidad de Sevilla).
R.~G.~J. acknowledges financial help from VPPI-US (Universidad de Sevilla).
We thank J.~M.~Ud\'ias and M.~V.~Ivanov for fruitful discussions on the RMF 
calculations.

\appendix
\section{Definition of the scaling functions}\label{apendiceScaling}

Within the context of the Relativistic Fermi Gas (RFG) model, the
scaling variable is defined as (see
\cite{Alberico88,Donnelly99a,Donnelly99b}) \ba \psi' \equiv
\frac{1}{\sqrt{\xi_F}}\frac{\lambda'-\tau'}
             {\sqrt{(1+\lambda')\tau'+\kappa\sqrt{\tau'(\tau'+1)}}}\,,
\ea where $\xi_F = \sqrt{1+(k_F/M)^2}-1$, $\kappa=q/(2M)$,
$\lambda'=\omega'/(2M)$ and $\tau=\kappa^2-\lambda'^2$. $M$ is the
nucleon mass and $k_F$ is the Fermi momentum~\cite{Maieron02}.
Additionally, we have introduced the variable
$\omega'=\omega-E_{shift}$. The quantity $E_{shift}$, which is
different for each target nucleus~\cite{Maieron02}, is introduced to
account phenomenologically for the shift observed in the QE peak
when the cross section is plotted as a function of $\omega$.
Trivially, if $E_{shift}=0$ one recovers the unshifted scaling
variable $\psi$.

\subsection{Electromagnetic scaling functions}

For $N=Z$ nuclei the isovector ($T=1$) and isoscalar ($T=0$) EM longitudinal, $L$,
and transverse, $T$, scaling functions are: \ba f^{T=1,0}_{L,T}
\equiv k_F \frac{R^{T=1,0}_{L,T}(\kappa,\lambda)}
                       {G^{T=1,0}_{L,T}(\kappa,\lambda)}\,.
\ea

We have introduced the elementary cross sections
\ba
G^{T=1,0}_{L,T}(\kappa,\lambda)
     =\frac{1}{2\kappa {\cal D}}U^{T=1,0}_{L,T}(\kappa,\lambda)\,,
\ea
where
\ba
U^{T=1,0}_L(\kappa,\lambda) = \frac{\kappa^2}{\tau}
             \left[H^{T=1,0}_E+W^{T=1,0}_2\Delta\right] \\
U^{T=1,0}_T(\kappa,\lambda) = 2\tau H^{T=1,0}_M+W^{T=1,0}_2\Delta\,,
\ea
with
\ba
H^{T=1,0}_{E,M} &=& \frac{Z + N}{4}(G_{E,M}^{T=1,0})^2 \label{H10}\\
W^{T=1,0}_2 &=& \frac{1}{1+\tau}\left[H^{T=1,0}_E+\tau H^{T=1,0}_M\right]\,.\label{W210}
\ea
$Z$ and $N$ are the number of protons and neutrons in the target nucleus, repectivaly.
Finally,
\ba
\Delta &\equiv& \xi_F(1-\psi^2)\left[\frac{\sqrt{\tau(\tau+1)}}{\kappa}
                + \frac{\xi_F}{3}\frac{\tau}{\kappa^2}(1-\psi^2)\right] \nonumber\\
                \\
{\cal D} &\equiv& 1 + \frac{1}{2}\xi_F(1+\psi^2)\,.
\ea
Note that Pauli-blocking effects have been neglected here.

Notice that we have introduced the isoscalar and isovector EM form
factors, $G_{E,M}^{T=1,0}$, which in terms of the more familiar
proton and neutron ones are \ba
G_{E,M}^{T=0} = G_{E,M}^{p} + G_{E,M}^{n} \\
G_{E,M}^{T=1} = G_{E,M}^{p} - G_{E,M}^{n}\,. \ea In this work, the
GKex VMD-based model~\cite{Lomon01,Lomon02,Crawford10} has been used
for the proton and neutron EM form factors.

The total longitudinal, $L$, and transverse, $T$, scaling functions
are defined as usual: \ba f_{L,T} \equiv k_F
\frac{R_{L,T}(\kappa,\lambda)}
                       {G_{L,T}(\kappa,\lambda)}\,,
\ea
where $G_{L,T}$ (and $U_{L,T}$) are built as above but
with the following
definition of $H_{E,M}$ and $W_2$:
\ba
H_{E,M} &=& Z(G_{E,M}^p)^2 + N(G_{E,M}^n)^2 \\
W_2 &=& \frac{1}{1+\tau}\left[H_E+\tau H_M\right]\,.
\ea

\subsection{Charge-changing neutrino and antineutrino scaling functions}

In this case the current is purely isovector ($T=1$).
As usual one defines
\ba
f^{\nu(\bar{\nu})}_{K} \equiv k_F \frac{R^{\nu(\bar{\nu})}_{K}(\kappa,\lambda)}
                       {G_{K}(\kappa,\lambda)}\,,
\ea
where $K=L,T,CC,CL,LL,T'$ for $VV$, $AA$ and $VA$ cases.
The elementary cross sections are
\ba
G_{K}(\kappa,\lambda)
     =\frac{1}{2\kappa {\cal D}}U_{K}(\kappa,\lambda)\,,\\
\ea
which are defined in terms of
\ba
U_L^{VV} &=& \frac{\kappa^2}{\tau}\left[H_E^{T=1} + W_2^{T=1}\Delta\right] \\
U_T^{VV} &=& 2\tau H_M^{T=1} + W_2^{T=1}\Delta
\ea
\ba
U_{CC}^{AA} &=& \frac{\kappa^2}{\tau}\left[\left(\frac{\lambda}{\kappa}\right)^2 H_A'
         + H_A\Delta\right] \\
U_{LL}^{AA} &=& \frac{\kappa^2}{\tau}\left[H_A'
         + \left(\frac{\lambda}{\kappa}\right)^2 H_A\Delta\right] \\
U_{CL}^{AA} &=& -\frac{\kappa^2}{\tau} \left(\frac{\lambda}{\kappa}\right)
          \left[H_A'+ H_A\Delta \right]
\ea
\ba
U_T^{AA} &=& H_A \left[2(1+\tau) + \Delta \right] \\
U_{T'}^{VA} &=& 2\sqrt{\tau(1+\tau)} H_{VA} \left[1 + \Delta'
\right]\,.\label{UVA} \ea
The functions $H_{E,M}^{T=1}$ are given in
Eq.~(\ref{H10}) but in this case the factor $(Z+N)$ should be
replaced by ${\cal N}$ which is $N$ or $Z$ for neutrino or
antineutrino charge-changing reactions. We have also introduced the
functions: \ba
H_A &=& {\cal N} \left[G_A^{T=1}\right]^2 \\
H_A' &=& {\cal N} \left[G_A'\right]^2 \\
H_{VA} &=& {\cal N} G_M^{T=1} G_A^{T=1}
\ea
with
\ba
&&G_A' \equiv G_A^{T=1} - \tau G_P^{T=1} = \frac{1}{1+|Q^2|/m_\pi^2}G_A^{T=1}\,. \nonumber \\
&& \ea and $G_A^{T=1} = g_A \left(1+|Q^2|/M_A^2\right)^{-2}$, being
$g_A=1.2695$, $m_\pi$ the pion mass and $M_A=1.03$ GeV.

Finally, the quantity $\Delta'$ which appears in Eq.~(\ref{UVA}) is defined as
\ba
\Delta' = \frac{1}{\kappa\sqrt{1+1/\tau}}\frac{1}{2}\xi_F(1-\psi^2)\,.
\ea
Note that Pauli-blocking effects have also been neglected here.

\section{Parameterization of the reference scaling functions}\label{appendixSG}
In this Appendix we summarize the parameterization of the reference scaling functions.
The skewed-Gumbel (sG) function is
\begin{eqnarray}
 f_{sG}=S(\nu_0;\psi)f_G(\psi_0,\sigma,\beta;\psi) \, ,
\end{eqnarray}
where
\begin{eqnarray}
 S(\nu_0;\psi) &=& \frac{2}{1+e^{\nu/\nu_0}} \\
 f_G(\psi_0,\sigma,\beta;\psi) &=&
               \frac{\beta}{\sigma}e^{\nu}\exp[-e^{\nu}] \\
 \nu &=& -\left(\frac{\psi-\psi_0}{\sigma}\right)\,.
\end{eqnarray}

\begin{figure}[htbp]
     \centering
         \includegraphics[width=0.3\textwidth,angle=270]{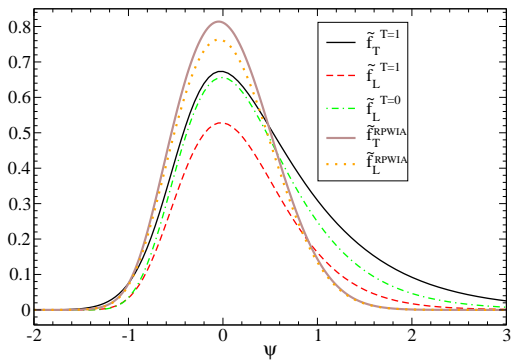}
     \caption{Reference scaling functions in SuSAv2 model.}
     \label{fig:patrones}
\end{figure}

In Table~\ref{table1} are shown the values of the free parameters
that fit the reference scaling functions $\tilde f_{L,T=1}$, $\tilde
f_{L,T=0}$ and $\tilde f_T$. In Fig.~\ref{fig:patrones} these {\it
reference} scaling functions are presented as functions of the
scaling variable $\psi$.

\begin{table}[htbp]
\centering
\begin{tabular}{c|ccc}
        &  $\tilde f_{L,T=1}$   & $\tilde f_{L,T=0}$ & $\tilde f_T$ \\
\hline
$\beta$     & $0.8923$   & $1.0361$  & $0.9425$ \\
$\sigma$    & $0.6572$   & $0.5817$  & $0.7573$ \\
$\psi_0$    & $0.1708$   & $0.02217$ & $-0.4675$ \\
$1/\nu_0$   & $-0.7501$  & $-0.1163$ & $2.9381$ \\
\hline
\end{tabular}
\caption{Values of the parameters that characterize the sG reference scaling functions.}
\label{table1}
\end{table}

The reference RPWIA scaling functions are
\begin{equation}
\tilde f_{L,T}^{RPWIA}
 = \frac{2(a_3)_{L,T}}{1+\exp\left({\frac{\psi-a_1}{a_2}}\right)}
\exp \left(-\frac{\left(\psi-a_4\right)^2}{a_5}\right)\,,
\end{equation}
with $a_1 = -0.892196$, $a_2 = 0.1792$, $(a_3)_L = 6070.85$,
$(a_3)_T = 6475.57$, $a_4 = 1.74049$, $a_5 = 0.64559$.

\section{Pauli Blocking in SuSA and SuSAv2}\label{appendixPB}

In this Appendix we show the effects of Pauli Blocking (PB) in the
SuSA and SuSAv2 models. The procedure employed to introduce PB in
the SuSA model was already presented in \cite{Megias14}. The
method, proposed in \cite{Rosenfelder:1980nd}, consists in
building a new scaling function by subtracting from the original
one, $f[\psi(\omega,q)]$, its ``mirror'' function,
$f[\psi(-\omega,q)]$ (see~\cite{Megias14} for details). In the RFG
this procedure yields exactly the same result as the usual way of
introducing Pauli blocking via theta-functions. However the method
can also be applied to models, like SuSA, which are not built
starting from a momentum distribution. The same procedure is used in
this work to introduce PB in SuSAv2 model.

\begin{figure}[htbp]
     \centering
         \includegraphics[width=0.48\textwidth,angle=0]{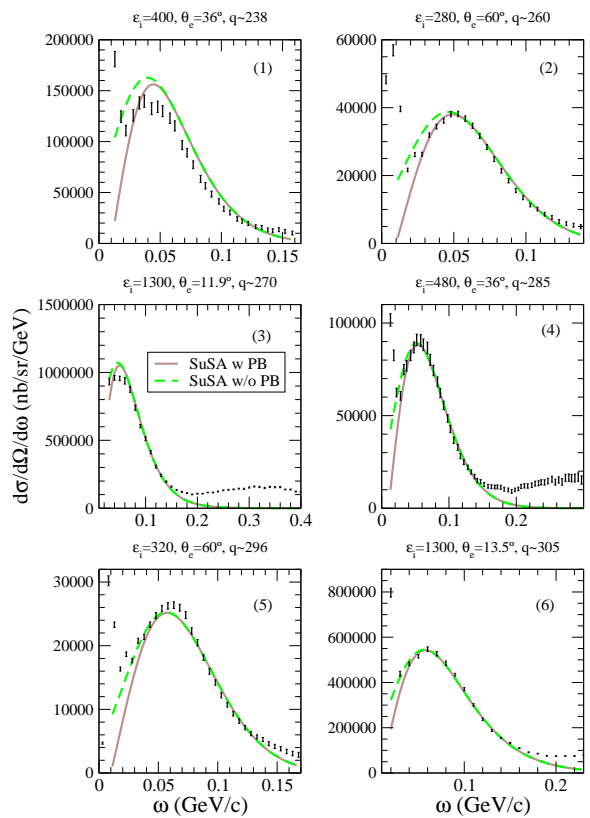}
     \caption{SuSA with and without Pauli Blocking is compared with data.
     $E_{shift}=10$ MeV has been employed.}
     \label{fig:SuSAPB}
\end{figure}

\begin{figure}[htbp]
     \centering
         \includegraphics[width=0.48\textwidth,angle=0]{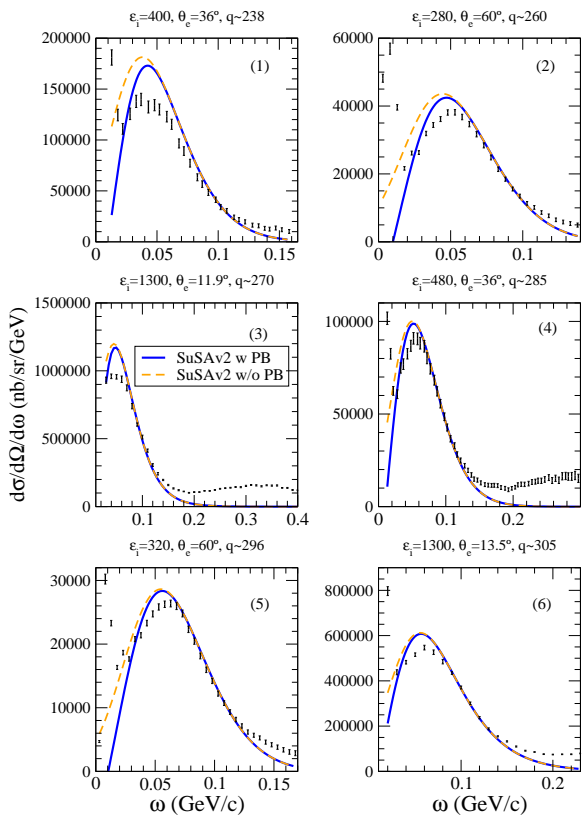}
     \caption{SuSAv2 with and without Pauli Blocking is compared with data.}
     \label{fig:SuSAv2PB}
\end{figure}

We comment on Fig.~\ref{fig:SuSAPB} where SuSA results with and
without PB are compared with a few sets of data at the kinematics in
which PB effects are significant, {\it i.e.,} very low $q$. In order
to fit the position of the peak better, in this case we have used a
shift energy of 10 MeV in the SuSA model (compared with the 20 MeV
used in Figs.~\ref{fig:lowerq_r}-\ref{fig:higherq_r}.
This makes the comparison
with data easier and allows us to focus on PB effects, namely, the
width and peak height of the cross sections. In general we conclude
that the agreement between SuSA and data improves when PB is
introduced. SuSA without PB (green-dashed) produces cross sections
too wide, while SuSA with PB (brown) provides narrower cross
sections in better agreement with data. This is particularly true
for instance in panels $(1)$ and $(2)$ in Fig.~\ref{fig:SuSAPB}. The
same comments apply to Fig.~\ref{fig:SuSAv2PB} where SuSAv2 with and
without PB is compared with the same set of low-$q$ data. The lowest
energy transfer data, corresponding to the excitation of resonant
and collective states, cannot be described by any of the present
models.

A clear difference between SuSA and SuSAv2 (Figs.~\ref{fig:SuSAPB}
and \ref{fig:SuSAv2PB}) is that the latter clearly overestimates the
data in the region below and close to the peak. However, in all
cases the maximum is placed at the region $\omega\lesssim50-60$ MeV
where, as discussed in Sect.~\ref{ee-results}, the validity of the
models based on IA is questionable and no definitive conclusions can
be drawn based on comparison of model and data in this
$\omega$-region.


\bibliographystyle{apsrev4-1}

\bibliography{bibliography}

\end{document}